\documentclass[preprint,12pt]{elsarticle}
\usepackage{amssymb}
\usepackage{amsmath}
\usepackage{subfigure}
\usepackage{float}
\usepackage{hyperref}
\usepackage{graphicx}
\usepackage{bm}
\usepackage{epstopdf}
\usepackage{lineno}

\usepackage{xcolor}
\newcommand{\reviewI}[1]{#1}
\newcommand{\reviewII}[1]{#1}
\newcommand{\rewrite}[1]{#1}
\newcommand{\Fix}[1]{#1}

\journal{Nuclear Instruments and Methods in Physics Research Section A}

\begin{document}

\begin{frontmatter}

\title{\boldmath Development of a simulation and analysis framework for N$\nu$DEx experiment}

\author[a,b]{Tianyu Liang}
\author[a,b]{Hulin Wang$^*$}
\author[a,b]{Dongliang Zhang}
\author[a,b]{Chaosong Gao}
\author[a,b]{Xiangming Sun}
\author[a,b]{Feng Liu}
\author[a,b]{Jun Liu}
\author[c]{Chengui Lu}
\author[c]{Yichen Yang}
\author[c]{Chengxin Zhao}
\author[c]{Hao Qiu}
\author[a,b]{Kai Chen$^{**}$}

\journal{Nuclear Instruments and Methods in Physics Research A}

\let\oldfootnoterule\footnoterule
\renewcommand{\footnoterule}{\oldfootnoterule\vspace{2em}}

\cortext[cor1]{Corresponding author: hulin.wang@ccnu.edu.cn}
\cortext[cor2]{Corresponding author: chenkai@ccnu.edu.cn}

\address[a]{PLAC, Key Laboratory of Quark \& Lepton Physics (MOE), 
            Central China Normal University, 
            152 Luoyu Rd., Wuhan 430000, China}

\address[b]{Hubei Provincial Engineering Research Center of Silicon Pixel Chip \& Detection Technology,
            Central China Normal University, 
            152 Luoyu Rd., Wuhan 430000, China}

\address[c]{Institute of Modern Physics, 
            Chinese Academy of Sciences, 
            509 Nanchang Rd., Lanzhou 730000, China}

\begin{abstract}
N$\nu$DEx aims to search for the neutrinoless double beta decay in $^{82}$Se using a high pressure $^{82}$SeF$_6$ gas time projection chamber (TPC). This paper presents a simulation and \reviewII{analysis} framework developed specifically for the N$\nu$DEx experiment.

Using density functional theory and two-temperature theory, the reduced mobilities of SeF$_5^-$ and SeF$_6^-$ ions in SeF$_6$ were calculated, yielding values of \rewrite{0.444 and 0.430} $\mathrm{cm^2V^{-1}s^{-1}}$ respectively, with an estimated \rewrite{uncertainty within 3\%}. 

The TPC geometry, featuring a cathode, focusing plane, and anode structure, was modeled in COMSOL to compute electric fields. Signal and background events were generated using BxDecay0 and Geant4, while Garfield++ was employed to simulate charge transport and signal induction. Three-dimensional tracks were reconstructed from drift-time differences \reviewII{between the two assumed ion species} using a breadth-first search algorithm.

\rewrite{To demonstrate the framework's analytical capability}, topological variables were \reviewII{taken} from reconstructed tracks and used to define selection criteria. A boosted decision tree was \reviewII{then} \rewrite{implemented to benchmark the signal-background separation}. \rewrite{This simulation framework successfully validates the complete experimental workflow, serving as a robust tool for detector design and future sensitivity studies in the N$\nu$DEx experiment.}

\end{abstract}

\end{frontmatter}

\section{Introduction}
\label{sec:intro}

The search for neutrinoless double beta decay ($0\nu\beta\beta$) in $\mathrm{^{82}Se}$ using a high-pressure $\mathrm{SeF_6}$ gas Time Projection Chamber (TPC) was first proposed by D.R. Nygren et al.~\cite{Nygren_2018}. This concept evolved into the $\mathrm{N\nu DEx}$ experiment~\cite{nvdexcdr2023}, which is planned for deployment at the China Jinping Underground Laboratory (CJPL)~\cite{Cheng:2017usi}. 
The $^{82}$Se features a Q value of 2.998~MeV and a natural abundance of 8.7\%.
\reviewI{The N$\nu$DEx TPC plans to operate at approximately 10 bar of $\mathrm{^{82}Se}$-enriched SeF$_6$ gas.
The first stage of the project, N$\nu$DEx-100, will utilize 100~kg $\mathrm{^{82}SeF_6}$ within a sensitive volume of 1.2 m$^3$ (at a gas density of 0.082 $\mathrm{g/cm^3}$).
Assuming a background-free analysis, it is projected to achieve a half-life sensitivity of $4 \times 10^{25}$ ($4 \times 10^{26}$) years for natural (enriched) gas after 5 years of operation.}

As a highly electronegative hexafluoride, SeF$_6$ \Fix{efficiently} captures free electrons to form negative ions. To avoid the signal fluctuations associated with ion stripping and avalanches, N$\nu$DEx adopts a direct charge collection scheme for these ionic carriers. This approach is crucial to achieving the target 1\% FWHM energy resolution.
To meet this requirement, a charge sensing and readout chip named Topmetal-S~\cite{Yang_2024}
\reviewII{, which is a member of the Topmetal series~\cite{Fan:2014hka},} 
has been developed. The readout plane of $\mathrm{N\nu DEx}$-100 comprises approximately 10,000 such pixelated sensing units~\cite{nvdexcdr2023}. To enhance charge collection efficiency, a focusing plane positioned above the anode guides drifting ions \reviewII{towards} the sensing area.

TPCs are established tools in $0\nu\beta\beta$ searches due to their superior \reviewII{tracking capabilities}, as demonstrated by experiments such as EXO~\cite{MAuger_2012}, nEXO~\cite{Adhikari_2022}, NEXT~\cite{next2023}, and PandaX-III~\cite{PandaX2017}.  In the hexafluoride gases, previous studies~\cite{Gary,Stockdale,BRION1969197} have shown that anions form through the attachment of free electrons to neutral molecules:
\begin{equation}
\label{FreeAttach}
    e^- + XF_n \rightarrow XF_n^{-*}
\end{equation}
where X represents S, Se, Te, or Mo. The excited anion can stabilize via collisions:
\begin{equation}
\label{De-excited}
    XF_n^{-*} \rightarrow XF_n^{-}
\end{equation}
or undergo dissociation:
\begin{equation}
\label{Disso}
    XF_n^{-*} \rightarrow XF_{n-1}^{-} + F
\end{equation}
Additionally, \reviewII{the} excited anion \reviewII{assumed} may also undergo auto-detachment:
\begin{equation}
\label{Det}
    XF_n^{-*}\rightarrow XF_n +e^-
\end{equation}

Ion mobility, K, defines the velocity $v$ of an ion drifting in a gas under under an electric field E:
\begin{equation}
    v = K E
\end{equation}
To account for variations in pressure and temperature, the reduced mobility K$_0$ is used:
\begin{equation}
    K_0 = K \frac{P}{P_0}\frac{T_0}{T}
    \label{eq:kk0}
\end{equation}
where P$_0$ = 1 atm and T$_0$ = 273.15 K. For an ion-TPC, characterizing the relationship between \reviewII{ion} drift velocity and the electric field is essential for accurate 3D track reconstruction.
Ions in a gas experience frequent collisions with surrounding molecules, \reviewII{thus their mobilities are linked} to the collision cross section. The two-temperature theory~\cite{GANDHI2023100191} describes the dynamics of these charged carriers in gases, in particular their energy exchange with the background medium. The first approximation, commonly known as the Mason-Schamp expression \reviewII{states}~\cite{Viehland1975GaseousLM}:
\begin{equation}
    K = \frac{3}{16}\frac{q}{n}(\frac{2 \pi}{\mu k_b T_{eff}})^{\frac{1}{2}}\frac{1}{\Omega}
    \label{MSeq}
\end{equation}
where q is the ion charge, n is the gas number density, $k_b$ is the Boltzmann constant, and $\mu = \frac{mM}{m+M}$ is the reduced mass (with $M$ and $m$ being the masses of the ion and gas molecule, respectively). The effective temperature, $T_{\text{eff}} = T + \frac{mv^2}{3k_b}$, accounts for the electric-field-induced kinetic energy. \reviewII{In the low electric field limit, $T_{\text{eff}} \approx T$}. The term $\Omega$ stands for the collision cross section. 

\begin{figure}
    \centering
    \includegraphics[width=0.9\linewidth]{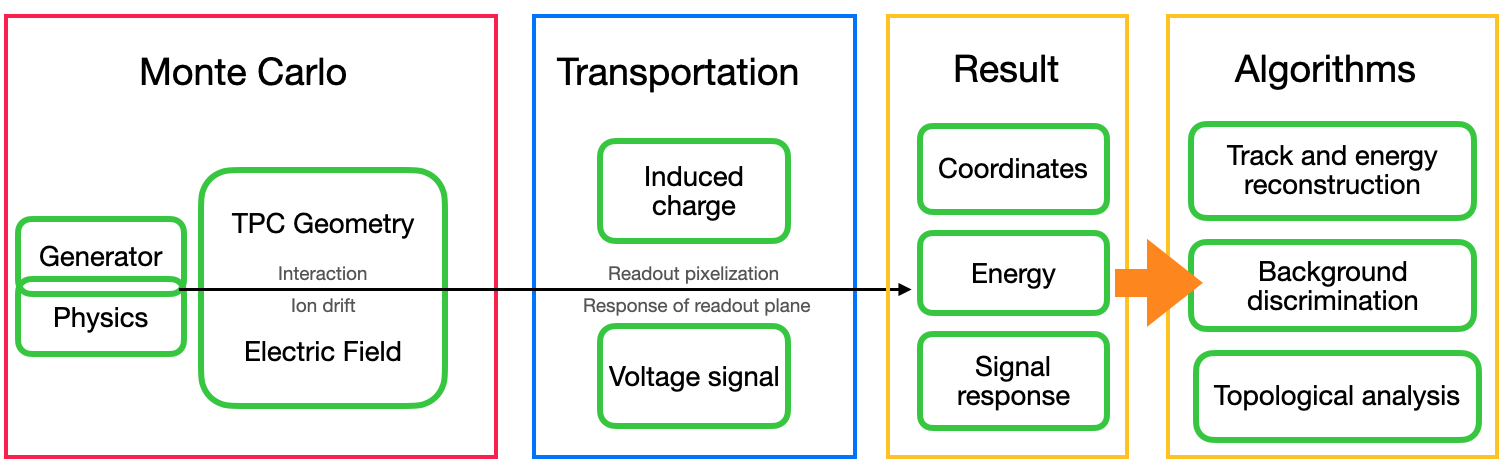}
    \caption{The simulation and analysis framework mainly consists of four components: event generation and physical interactions, charge transportation and signal inductions, detector response, and analysis algorithms.}
    \label{fig:Framework}
\end{figure}

The overall simulation and analysis framework developed for this work is illustrated in Fig.~\ref{fig:Framework}.
Section~\ref{sec:Calculation} details the investigation of reduced mobilities for the ion charge carriers. 
Section~\ref{sec:sim} describes the simulation architecture, covering event generation, physical interactions, charge transportation, signal inductions, and detector responses.
Finally, Section~\ref{sec:analy} presents the algorithms for track reconstruction and background discrimination.

\section{Ion mobility in \texorpdfstring{$\mathrm{SeF_6}$}{SeF6}}
\label{sec:Calculation}
\subsection{Candidate ion species in \texorpdfstring{$\mathrm{SeF_6}$}{SeF6}}

Although anion yields in high-pressure pure $\mathrm{SeF_6}$ have not been directly measured, prior studies offer valuable insights. Low-pressure investigations~\cite{Gary, BRION1969197} predominantly observe the $\mathrm{SeF_5^-}$ species. However, reducing the number density through admixtures of $\mathrm{SF_6}$ or nitrogen enables the detection of both $\mathrm{SeF_6^-}$ and $\mathrm{SeF_5^-}$. 
Nygren’s proposal~\cite{Nygren_2018} postulated that a stability mechanism for \reviewII{the excited} $\mathrm{SeF_6^-}$ \reviewII{ion} might exist, potentially similar to that of $\mathrm{SF_6^-}$. However, the specific dissociation mechanism and electron attachment properties were not explicitly modeled, and the lifetime of $\mathrm{SeF_6^-}$ at high pressure remains unverified. 

Nevertheless, the formation of $\mathrm{SeF_6^-}$ cannot be entirely excluded, particularly under specific admixture or pressure conditions. In this study, we treat both $\mathrm{SeF_5^-}$ and $\mathrm{SeF_6^-}$ \reviewII{negative ions} as candidate charge carriers, adopting a 20:80 yield ratio~\cite{Nygren_2018,Gary} to evaluate \reviewII{track} reconstruction performance under a dual-ion hypothesis.
If future measurements confirm the absence of $\mathrm{SeF_6^-}$ under experimental conditions, the framework can be simplified to single-species model, although the absolute z-position reconstruction would then require alternative methods. 

\subsection{Theoretical calculation and validation}

\subsubsection{Molecular modeling and optimization}

Ion mobility depends on molecular geometry, gas density, temperature, electric field strength, \Fix{polarizability}, and the ion–gas mass ratio. Initial atomic coordinates were obtained using Avogadro~\cite{Avogadro}, which employs the Universal Force Field (UFF) to reproduce structural features across the periodic table. 
The Avogadro output files were further refined with Gaussian16 (G16)~\cite{g16} to simulate the electronic structure. Calculations were performed using the B3LYP functional~\cite{Becke,Lee} with the 6-31G(d) basis set, which is commonly employed for structural optimization in density functional theory. The molecular charge was set to –1. Spin multiplicity, defined as $e_{\alpha} - e_{\beta} + 1$ for a given spatial electronic wave function, represents the number of possible spin orientations. For these negative ion, it corresponds to the number of unpaired electrons plus one.

Additionally, the polarizability was calculated. Although neutral $\mathrm{SeF_6}$ possesses an octahedral geometry with no permanent dipole moment, ion–molecule interactions induce a dipole that significantly influences transport. Polarizability is defined as:
\begin{equation}
    \alpha = \frac{|P|}{|E|}
\end{equation}
where P is the induced dipole moment and E is the electric field. The potential U between an ion and an induced dipole is expressed as: 
\begin{equation}
    U = -\frac{Z^2\alpha}{r^4}
    \label{InducedPotential}
\end{equation}
where Z is the charge \reviewII{of the ion}. Because this interaction scales with $r^{-4}$, it is referred to as the $4$–$\infty$ potential.
Using G16 (B3LYP/6-31G(d,p)), the polarizability tensor was obtained as:
\begin{equation*}
    \alpha  = \left(
    \begin{array}{ccc}
       35.085  &0 &0  \\
        0 & 35.084 &0 \\
        0&0&35.083
    \end{array}
    \right)
\end{equation*}
corresponding to an isotropic polarizability of 35.084 $a_0^3$ (or 5.206~$\mathring{\mathrm{A}}^3$), where $a_0$ is Bohr radius.

\subsubsection{Calculation of reduced mobility}

IMoS (Ion Mobility Software)~\cite{IMoS,COOTS2020105570} was used to calculate the mobility and collision cross section of ions in gas based on Eq.~\ref{MSeq}. The Trajectory-Diffuse Hard Sphere Scattering (TDHSS) method was employed, accounting for the ion–induced dipole interaction as described in Eq.~\ref{InducedPotential}.
While this potential captures long-range interactions, it omits short-range Pauli repulsion and the electronic cloud overlap. \reviewII{However, comparisons} with the full Lennard–Jones potential suggest that neglecting the repulsive term introduces an error less than 10\% under low field conditions (E/N $<$ 1 Td)~\cite{doi:https://doi.org/10.1002/3527602852.ch5b}.

\begin{figure}
    \centering
    \includegraphics[width=0.48\linewidth]{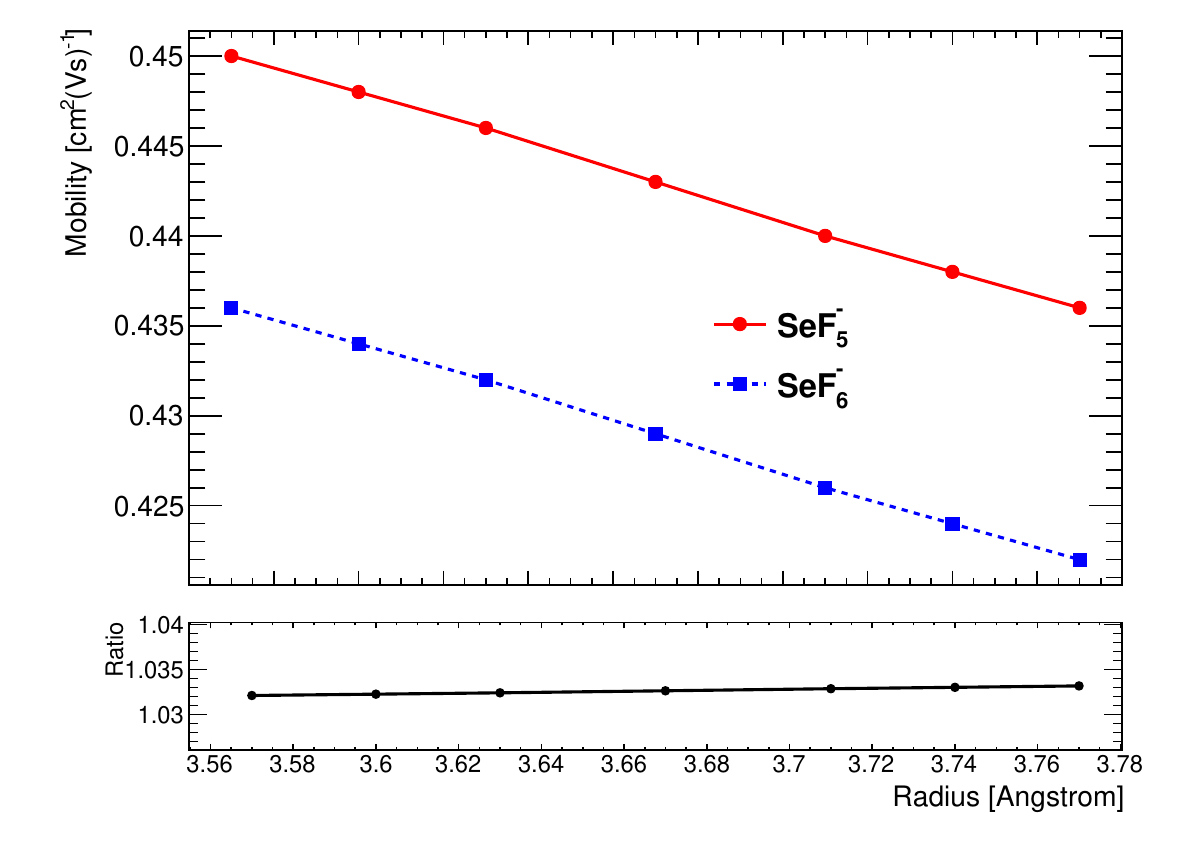}
    \includegraphics[width=0.48\linewidth]{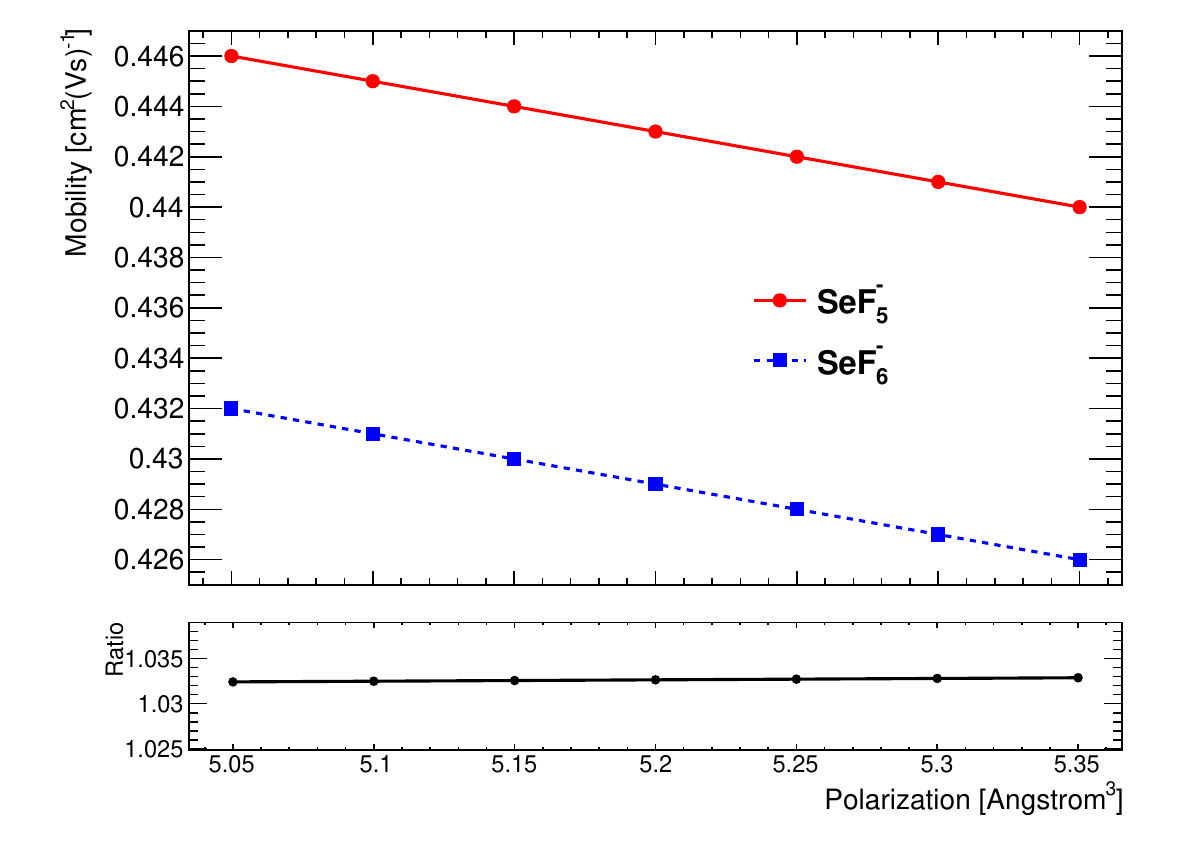}
    \caption{Sensitivity analysis of the calculated ion mobilities and their ratios. The dependence on gas molecular radius (left) and polarization (right) shows that while absolute mobility values vary, the mobility ratio ($\mathrm{K_{SeF_5^-}/K_{SeF_6^-}}$) remains stable. }
    \label{fig:senana}
\end{figure}

Using the optimized geometries and polarizabilities, we obtained reduced mobilities for $\mathrm{SeF_5^-}$ and $\mathrm{SeF_6^-}$ in $\mathrm{SeF_6}$ gas of 0.444 and 0.430 $\mathrm{cm^2 (Vs)^{-1}}$, respectively. The corresponding collision cross section are approximately 263.61 and 265.19 $\mathrm{\mathring{A}^2}$.
\rewrite{A sensitivity analysis was performed by varying the input gas molecular radius and polarization by a conservative $\pm 3\%$. As shown in Fig.~\ref{fig:senana}, while the absolute mobilities exhibit sensitivity to these parameter changes (shifting by $\pm$ 1.6\% for radius and $\pm$ 0.6\% for polarization), the mobility ratio ($\mathrm{K_{SeF_5^-} / K_{SeF_6^-}}$) remains stable at around 1.0326 (varying by less than 0.018\%), confirming the robustness of the dual-ion discrimination.}

To verify the results, reduced mobilities of $\mathrm{SF_6^-}$ and $\mathrm{SF_5^-}$ \reviewII{ions} in $\mathrm{SF_6}$ gas were also computed and compared to the experimental results, presented in Table~\ref{tab:Comparison}. \rewrite{A relative error of approximately 3\% was observed, providing a benchmark for the accuracy of our $\mathrm{SeF_6}$ calculations.}

\begin{table}[ht]
    \centering
    \begin{tabular}{c|c|c}
        \hline
        Ion & Calculation  &\reviewII{Experimental result~\cite{MARQUES2022166416}} \\
        \hline 
        SF$_6^-$ & $0.557$ &  $0.540 \pm 0.020$\\
        SF$_5^-$ & $0.580$ &  $0.599 \pm 0.022$\\
        \hline
    \end{tabular}
    \caption{Comparison between calculated and experimental reduced mobilities for $\mathrm{SF_6^-}$ and $\mathrm{SF_5^-}$ in $\mathrm{SF_6}$ gas.}
    \label{tab:Comparison}
\end{table}

\section{Simulation framework}
\label{sec:sim}

\begin{figure}
    \centering
    \includegraphics[width=0.46\linewidth]{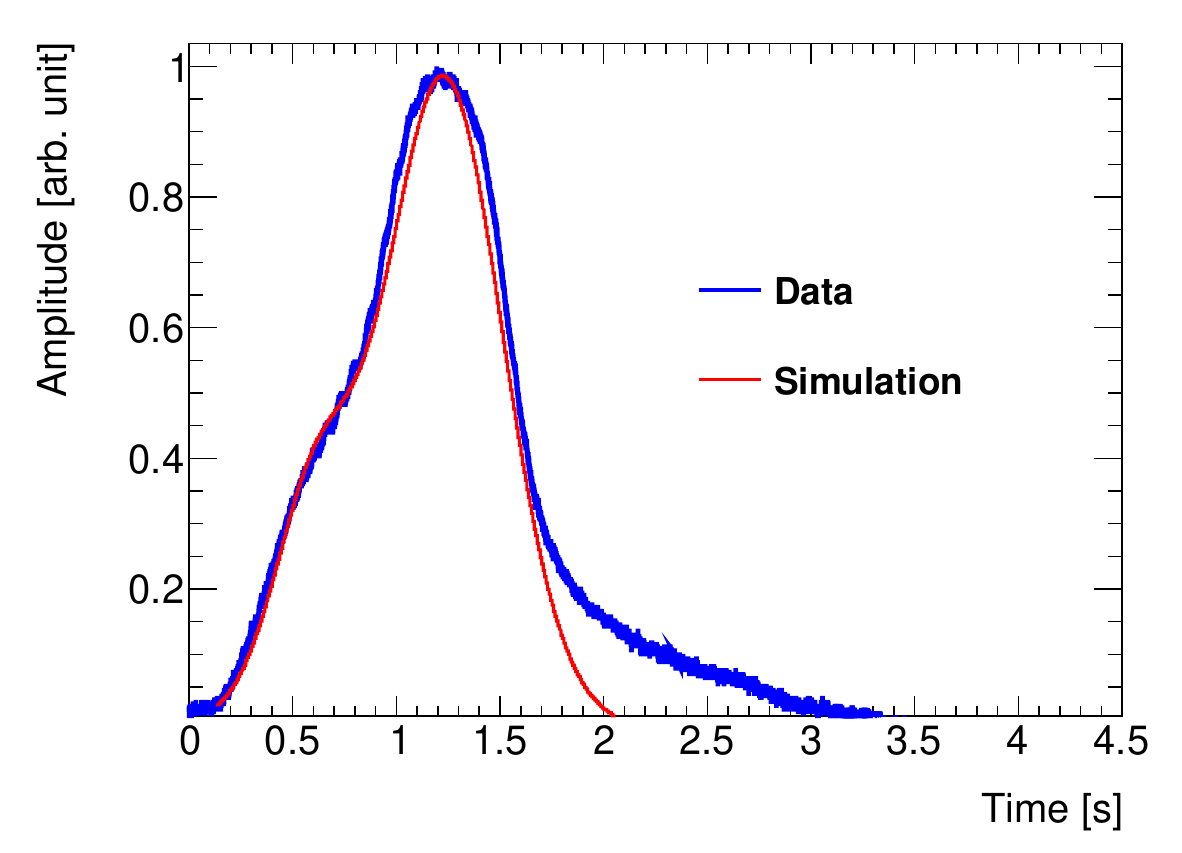}
    \includegraphics[width=0.52\linewidth]{ 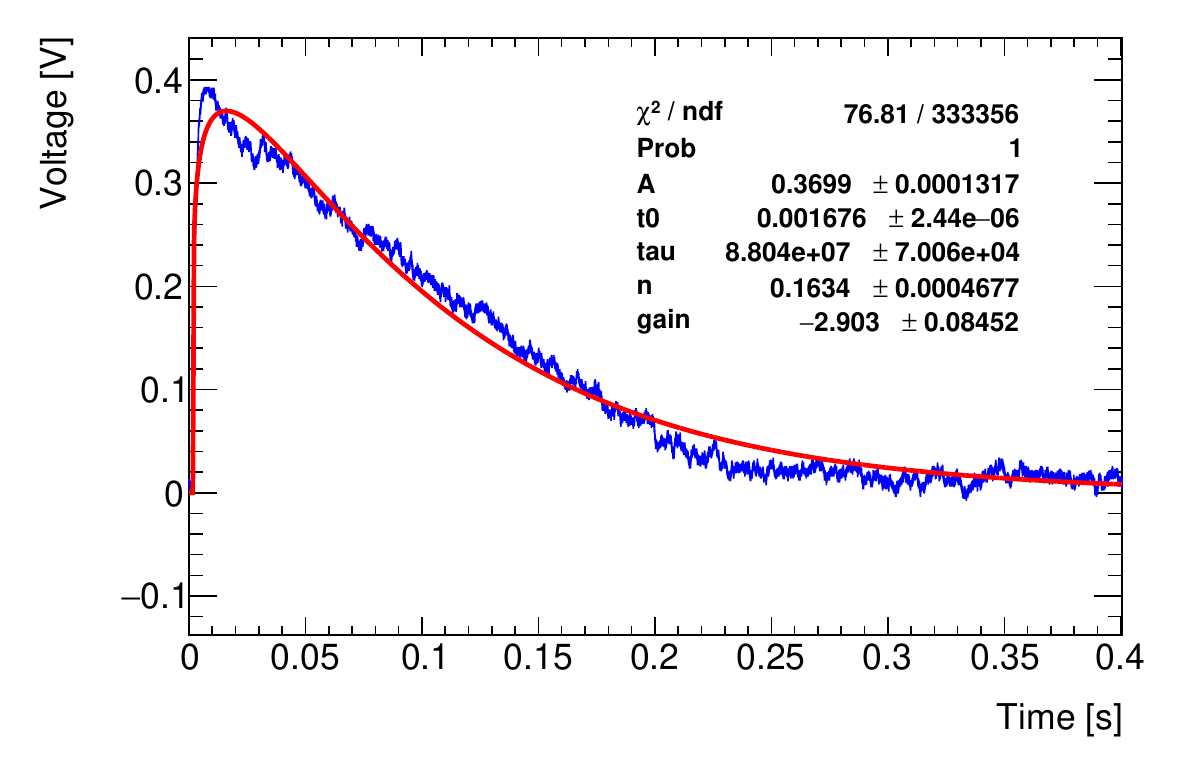}
    \caption{Left: \reviewII{comparison of simulated and experimentally measured waveforms for charge carriers SF$_5^-$ and SF$_6^-$ in SF$_6$ gas from Ref.~\cite{universe11050163}} as a function of the time of arrival; Right: output voltage waveform of the Topmetal-S chip from a test pulse, \rewrite{fitted with the induced signal generated by the framework convolved with the unipolar shaping function. The fit minimization was performed using the least squares method, and the reported $\chi^2$ represents the unweighted sum of squared residuals.} }
    \label{fig:WavCom}
\end{figure}

\rewrite{To ensure the reliability of the physical and electronic models integrated into our framework, we conducted two independent validations, as illustrated in Fig~\ref{fig:WavCom}}.
The left plot compares the simulated drift behavior of $\mathrm{SF_5^-}$ and $\mathrm{SF_6^-}$ ions in $\mathrm{SF_6}$ with \reviewII{experimental data} \reviewII{from} \rewrite{our previous work}~\cite{universe11050163} at a \reviewII{reduced} electric field of \reviewII{4.58~Td}. \reviewII{The simulation successfully reproduces the distinct arrival times of the two species,} after accounting for the uncertainties discussed in Ref.~\cite{universe11050163}. In the experimental waveform, the presence of heavier ions produces a long tail and a broader time distribution \reviewII{from} $\mathrm{SF_6^-}$, reflecting the delayed arrival of the slower species.

The right panel \rewrite{displays the measured output voltage waveforms of the Topmetal-S chip obtained using test pulses, as reported in Ref.~\cite{Liang_2024}. These are compared with \Fix{the fitting to the convolution of} the induced signal with the unipolar shaping function defined in the Garfield++ user guide~\cite{schindler2018garfield++}:
\begin{equation}
    f(t)  =g e^n (\frac{t}{t_p})^n e^{\frac{-t}{\tau}} , t_p  =n\tau
    \label{eq:Transfer}
\end{equation}
where n is the shaper order, $\tau$ is the shaping time constant, and g is the gain factor. Note that n was treated as a non-integer parameter to achieve a better fit. }
\reviewI{
Because the Topmetal-S chip exhibits a complex CMOS response that deviates from ideal shaping, the unipolar shaping function serves as an approximation for shaping the induced current in our simulations.
}

\begin{figure}
    \centering
    \includegraphics[width=0.8\linewidth]{ 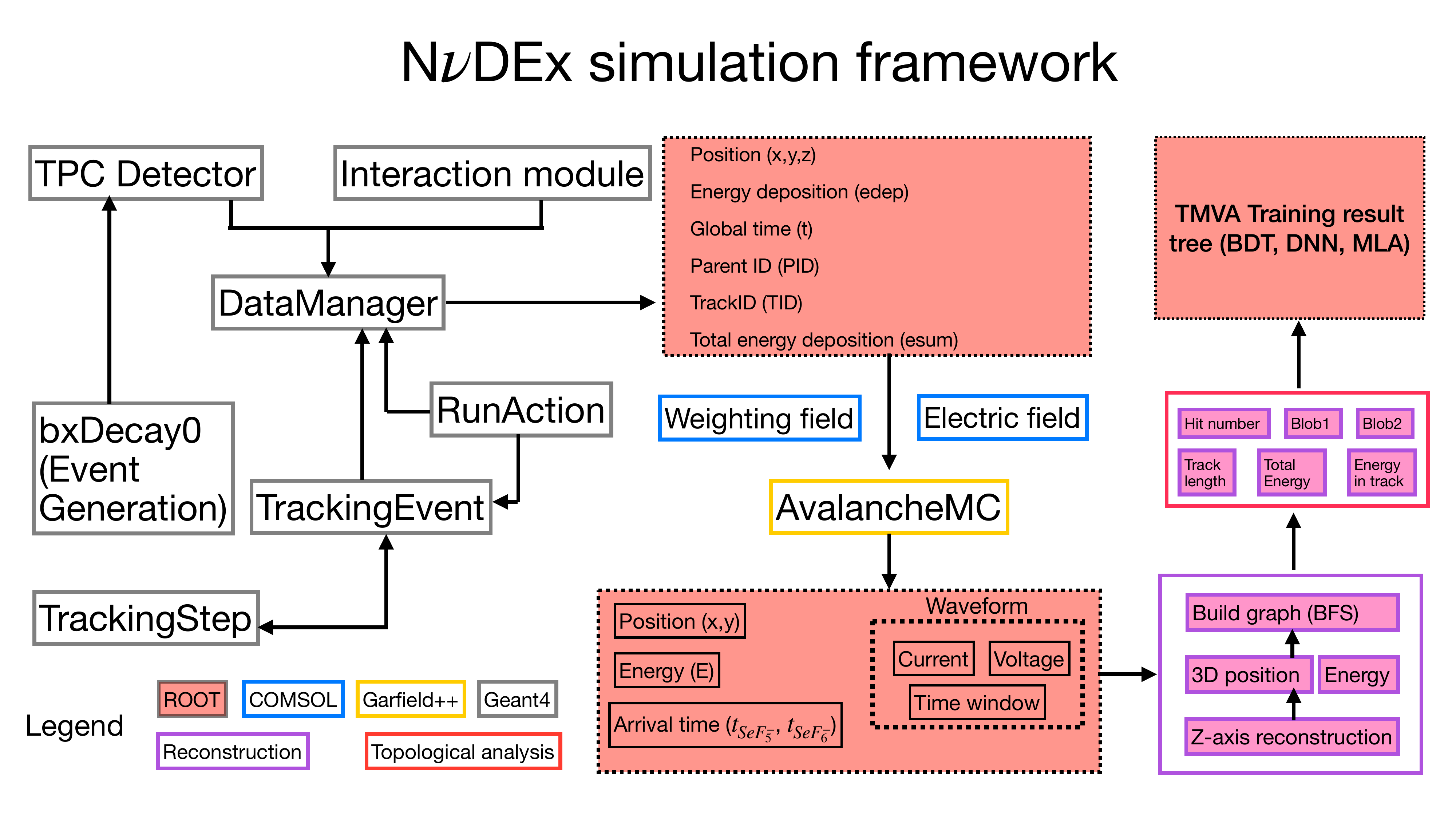}
    \caption{\rewrite{Detailed data flow diagram of the N$\nu$DEx simulation and analysis framework. The system orchestrates data exchange between Geant4 (particle interaction), COMSOL (field calculation), and Garfield++ (signal induction) via a central DataManager. Raw simulation data are stored in ROOT trees, while the downstream pipeline handles waveform digitization, 3D track reconstruction (BFS), and topological feature extraction for for TMVA classifiers.}}
    \label{fig:ArcDiag}
\end{figure}

\rewrite{The detailed software architecture of the N$\nu$DEx simulation framework are illustrated in Fig.~\ref{fig:ArcDiag}. The framework integrates four core external libraries: Geant4~\cite{AGOSTINELLI2003250,1610988,allison2016geant4}, Garfield++, COMSOL~\cite{COMSOL}, and ROOT~\cite{root_software,brun1997root}, orchestrated through a central data management system. }
\rewrite{First, the Event Generation utilizes BxDecay0~\cite{tretyak2023bxdecay0} and Geant4 to simulate primary particle decays and their interactions with the detector material. A dedicated DataManager module interfaces with RunAction, TrackingEvent, and TrackingStep classes in the framework \Fix{that} extracts the \Fix{true} information, which includes 3D positions $(x, y, z)$, energy deposition ($e_{dep}$), global time ($t$), and particle IDs.} 
\rewrite{Then, The detector signal response is processed using the AvalancheMC class in Garfield++, which imports electric and weighting fields from COMSOL to model charge drift and induction. The resulting signal response is digitized into current and voltage waveforms, recording arrival times and pixel hit coordinates.}
\rewrite{Finally, the Breadth-First Search (BFS) algorithm is employed to reconstruct 3D tracks. Topological features of reconstructed tracks, such as track length and blob energies, are extracted and fed into TMVA for multivariate signal-background classification.}

\subsection{Detector geometry structure}
\label{sec:Geo}

\begin{figure}
    \centering
    \includegraphics[width=0.8\linewidth]{ 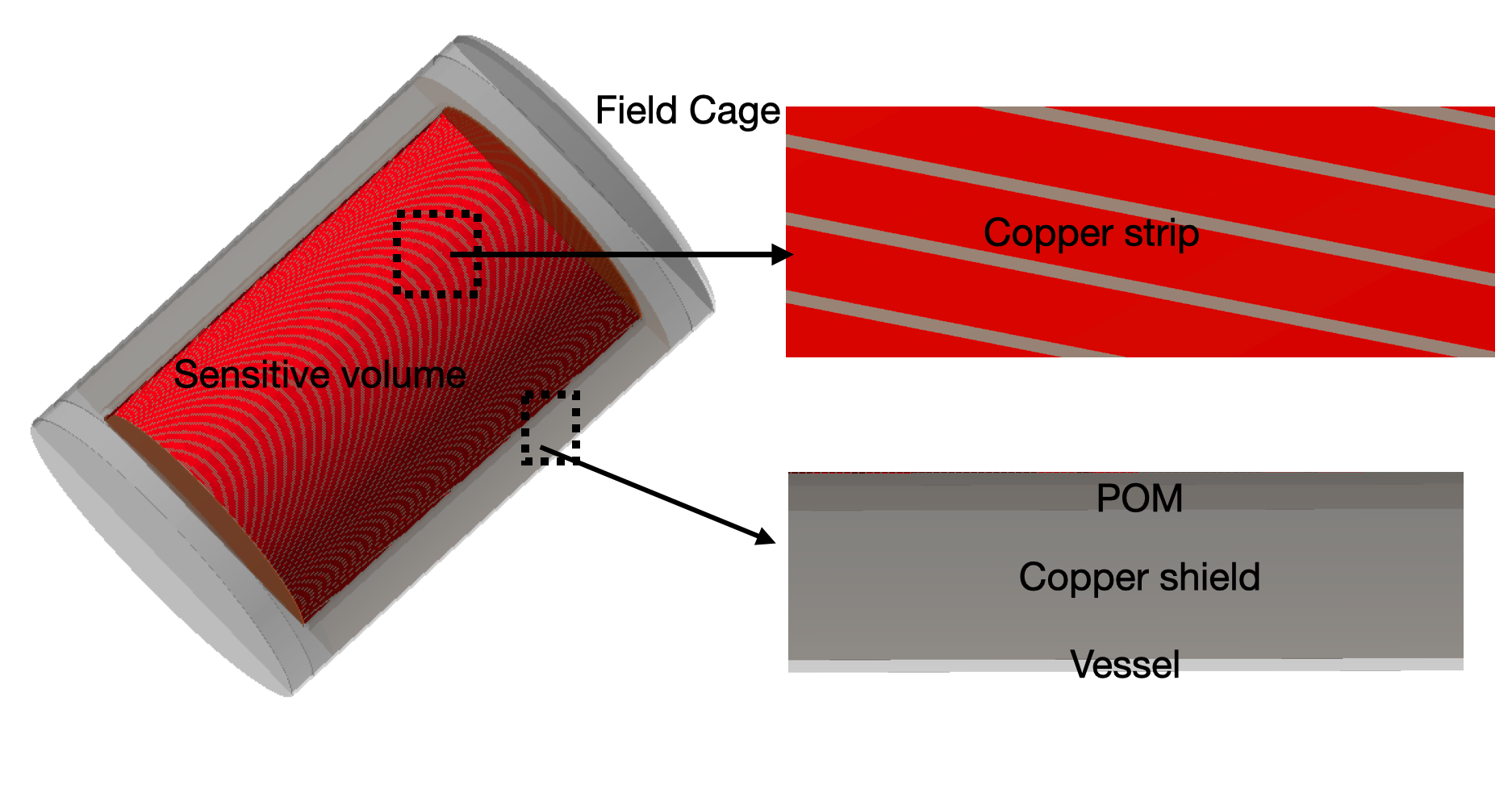}
    \caption{Geant4 model of the TPC. Layers \Fix{starting from inside} are sensitive volume, field cage, copper shield, and steel pressure vessel, respectively. }
    \label{fig:G4model}
\end{figure}

\rewrite{The basic geometry structure modeled in Geant4 consists of the pressure vessel, copper shield, field cage, and sensitive volume, as shown in Fig. \ref{fig:G4model}. The parameters of the model are detailed in Tab.~\ref{tab:geo}. 
The TPC is filled with $^{82}$SeF$_6$ gas at 10 atm and 20 $^\circ$C. 
Primary vertices of 0$\nu \beta\beta$ and 2$\nu \beta\beta$ events are distributed uniformly throughout the sensitive volume, while $\gamma$ backgrounds originate from the copper shield in this study.}

\begin{table}[h]
	\centering
	\begin{tabular}{c|c|c|c}
	\hline
	TPC layer & Radius [m] & Length [m] & Material\\
	\hline
	Sensitive volume & 0.5 & 1.6 & $^{82}$SeF$_6$ \\
    Field cage  & 0.503 & 1.6 & Copper + POM \\
	Copper shield + endcap  & 0.62 &  1.61 & Copper \\
	Steel pressure vessel  & 0.63 &2.0& Steel \\
	\hline
	\end{tabular}
	\caption{Geometrical parameters and material composition of the N$\nu$DEx-100 TPC modeled in Geant4.}
	\label{tab:geo}
\end{table}

\begin{figure}
\centering
\includegraphics[width=0.45\linewidth]{ 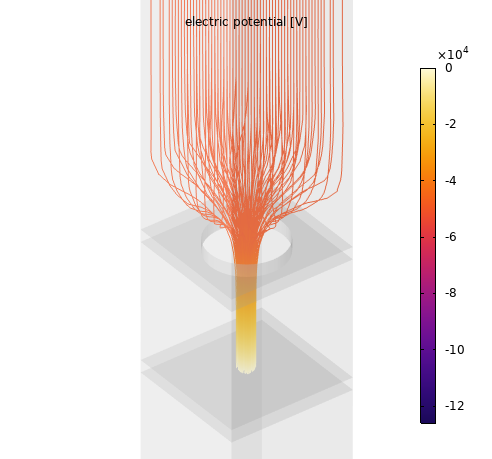}
\caption{COMSOL simulation of the readout unit cell. The electric field lines (streamlines) and \reviewII{electric potential} (color \reviewII{scale}) illustrate how \rewrite{the focusing hole funnels field lines towards the central sensing electrode.}}
\label{fig:Decmodel}
\end{figure}

\rewrite{The readout array are modeled by COMSOL, \Fix{and consist of} a focusing plane and an anode with approximately 10,000 Topmetal-S chips~\cite{nvdexcdr2023}. 
Each chip features a hexagonal charge collection electrode of 1 mm in diameter. Figure~\ref{fig:Decmodel} shows a unit cell with a 3 mm radius hole and a 1 cm pixel pitch.}

\begin{figure}
    \centering
    \includegraphics[width=0.5\linewidth]{ 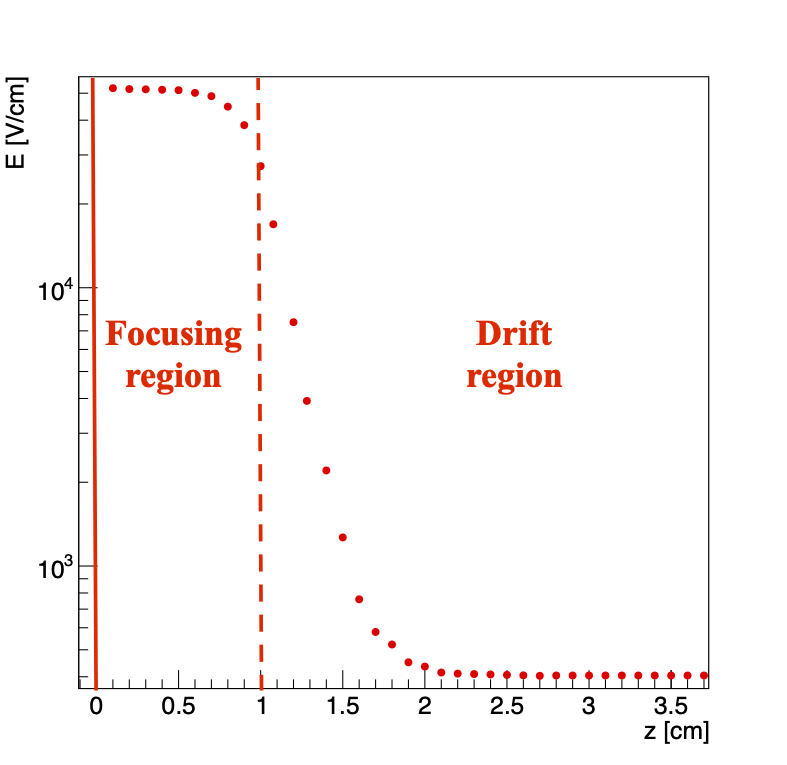}
    \caption{The electric field strength along the z-axis through the center of the focusing hole \Fix{as shown in Fig~\ref{fig:Decmodel}}.}
    \label{fig:zVsEF}
\end{figure}

The distance between the focusing plane and anode is set to 10 mm, while the distance between cathode and focusing plane is set to \rewrite{1600 mm}. The voltage of cathode and focusing plane are set to \rewrite{–114,000 V} and –50,000 V respectively, and the anode is grounded. The resulting electric field \Fix{in the} drift region is approximately 400 V/cm. The focusing-hole geometry distorts the electric field in the vicinity of the hole.
As shown in Fig.~\ref{fig:zVsEF}, along the vertical line through the center of the hole, the electric field strength begins to deviate as the distance to the \Fix{focusing} plane is less than 2 cm. In the focusing region, the field rises to approximately 50,000 V/cm, yielding a focusing-to-drift field ratio of roughly \rewrite{125}. \reviewII{This ensures a 100\% charge collection efficiency.}

\subsection{Energy deposition of electrons}
\label{sec:G4}

\begin{figure} [ht]
    \centering
    \includegraphics[width=0.8\linewidth]{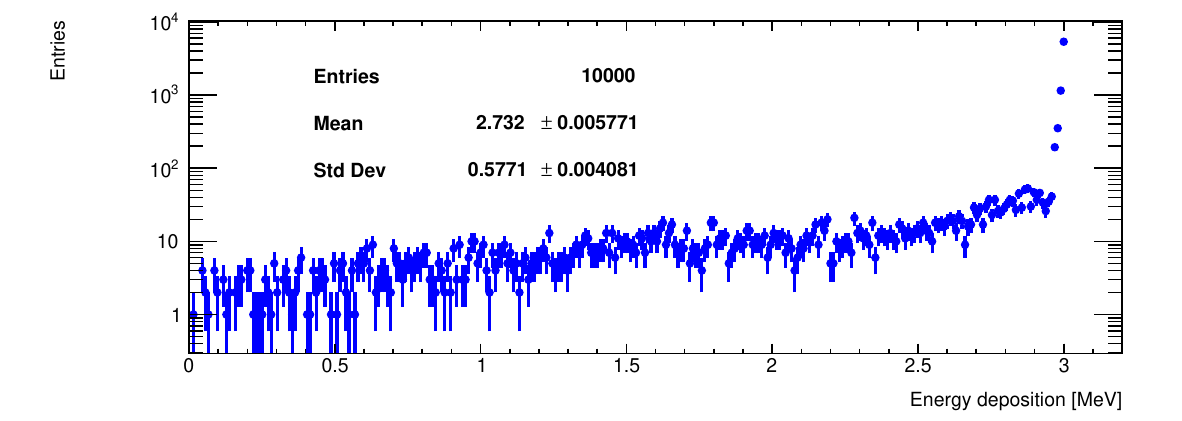}
    \caption{Simulated total energy deposition spectrum for $^{82}$Se $0\nu\beta\beta$ events in 10 atm SeF$_6$.}
    \label{fig:Edep}
\end{figure}

The BxDecay0, a C++ port of the Decay0/GENBB Monte Carlo code, is used to generate the standard decay and $0\nu \beta \beta$ processes for various radioactive nuclide\reviewII{s} of interest. For $0\nu \beta \beta$, the standard light Majorana neutrino mass mechanism is used. Particle interactions with detector materials are simulated using Geant4. \rewrite{The FTFP\_BERT physics list is used, with the electromagnetic module replaced by G4EmStandardPhysics\_option4~\cite{allison2016geant4} for accurate low-energy electron transport.}
Figure~\ref{fig:Edep} shows the distribution of total energy deposition inside the sensitive volume for \rewrite{10,000} simulated events. 
The sharp peak near 3 MeV corresponds to fully contained $0\nu \beta \beta$ events, 
while the tail results from $\gamma$-ray escape or events near the sensitive volume boundary.

\begin{figure}
    \centering
    \includegraphics[width=0.9\linewidth]{ 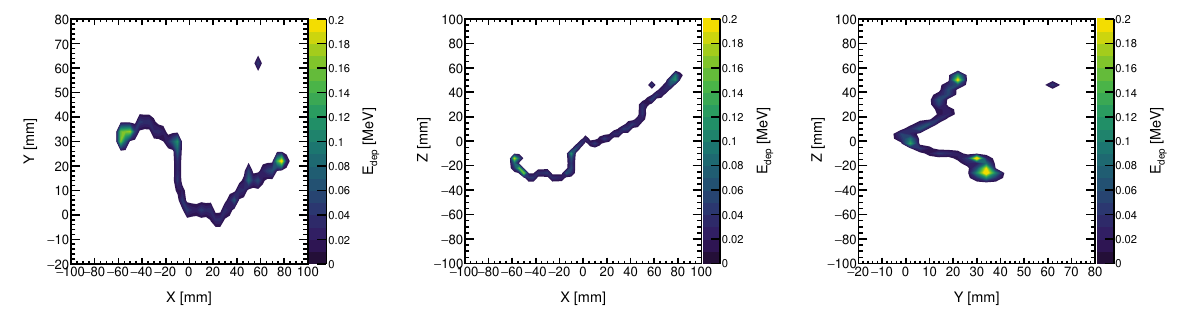}
    \includegraphics[width=0.9\linewidth]{ 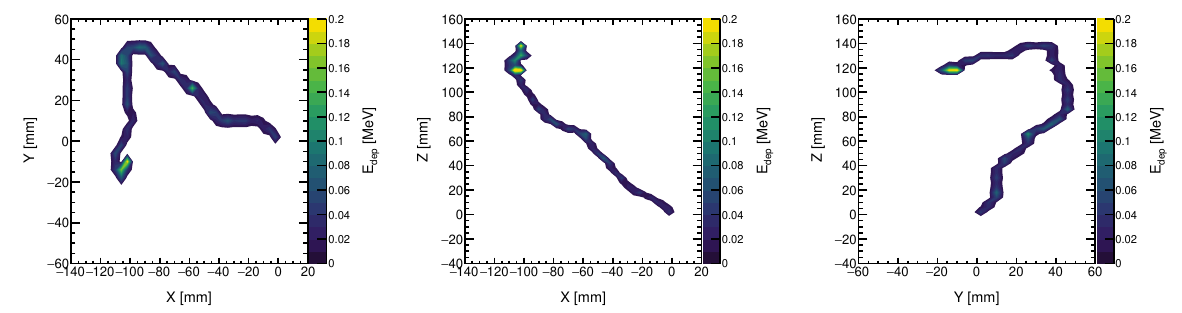}
    \caption{Projected 3D tracks for $0\nu \beta\beta$ (top) and single-electron background, both with total kinetic energy of \rewrite{2.998} MeV (bottom). }
    \label{fig:Proj}
\end{figure}

Figure~\ref{fig:Proj} shows projected 3D tracks for $0\nu \beta\beta$ and single-electron background, both with total kinetic energy of \rewrite{2.998} MeV. The “blobs” at track endpoints correspond to Bragg peaks. The $0\nu\beta\beta$ events typically exhibit two such blobs, whereas single-electron backgrounds show only one.

\subsection{\reviewII{Drift} of charge carriers}
\label{sec:Trans}
\rewrite{
The current induced by moving ions is described by the Shockley–Ramo theorem~\cite{Shockley1938CurrentsTC,RamoCurrent}:
\begin{equation}
    i = E_vqv
\end{equation}
where q is the charge of moving particle, $v$ is its instantaneous velocity, and $E_v$ is the component of the weighting field in the direction of $v$ at the charge's instantaneous position. 
}
The charge transport is simulated in Garfield++ using Monte Carlo integration (AvalancheMC) method.
Since specific data for $^{82}$SeF$_6$ is not \reviewII{available}, we adopt the W-value (35.8 eV) and Fano factor (0.28) of SF$_6$.
The ion diffusion coefficients are calculated using Einstein relation.
\reviewI{The gas is set at 293.15 K and 10 atm, and the ion mobilities for both SeF$_5^-$ and SeF$_6^-$ are derived from their reduced mobilities according to Eq.~\ref{eq:kk0}. Under an average electric field of 412 V/cm, the drift velocities for SeF$_5^-$ and SeF$_6^-$ are 19.6 cm/s and 19.0 cm/s, respectively. }

\begin{figure}[ht]
    \centering
    \includegraphics[width=0.5\linewidth]{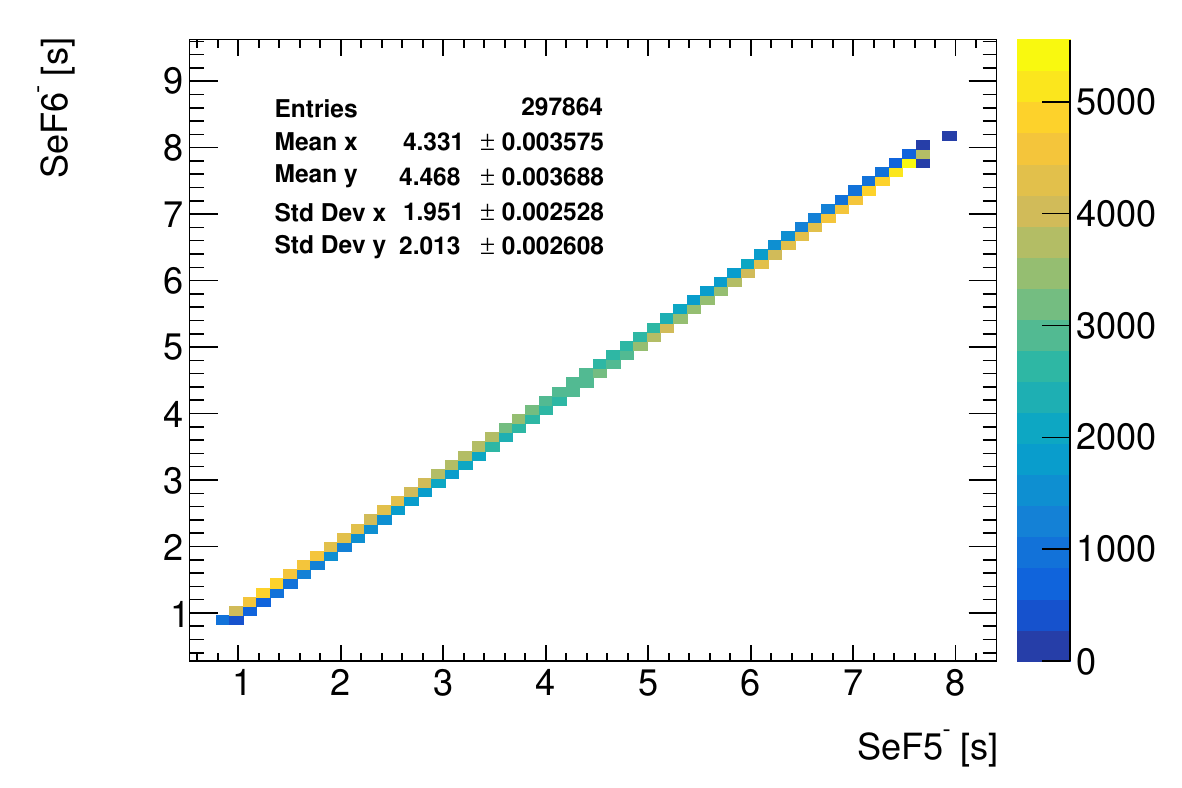}
    \caption{\rewrite{Simulated drift time distribution for $\mathrm{SeF_5^-}$ (x-axis) versus $\mathrm{SeF_6^-}$ (y-axis) at the anode. The color scale indicates the number of entries. The linear relationship confirms that drift time differences are consistent with the mobility ratio.}}
    \label{fig:TimeDis}
\end{figure}

\begin{figure}
    \centering
    \includegraphics[width=0.48\linewidth]{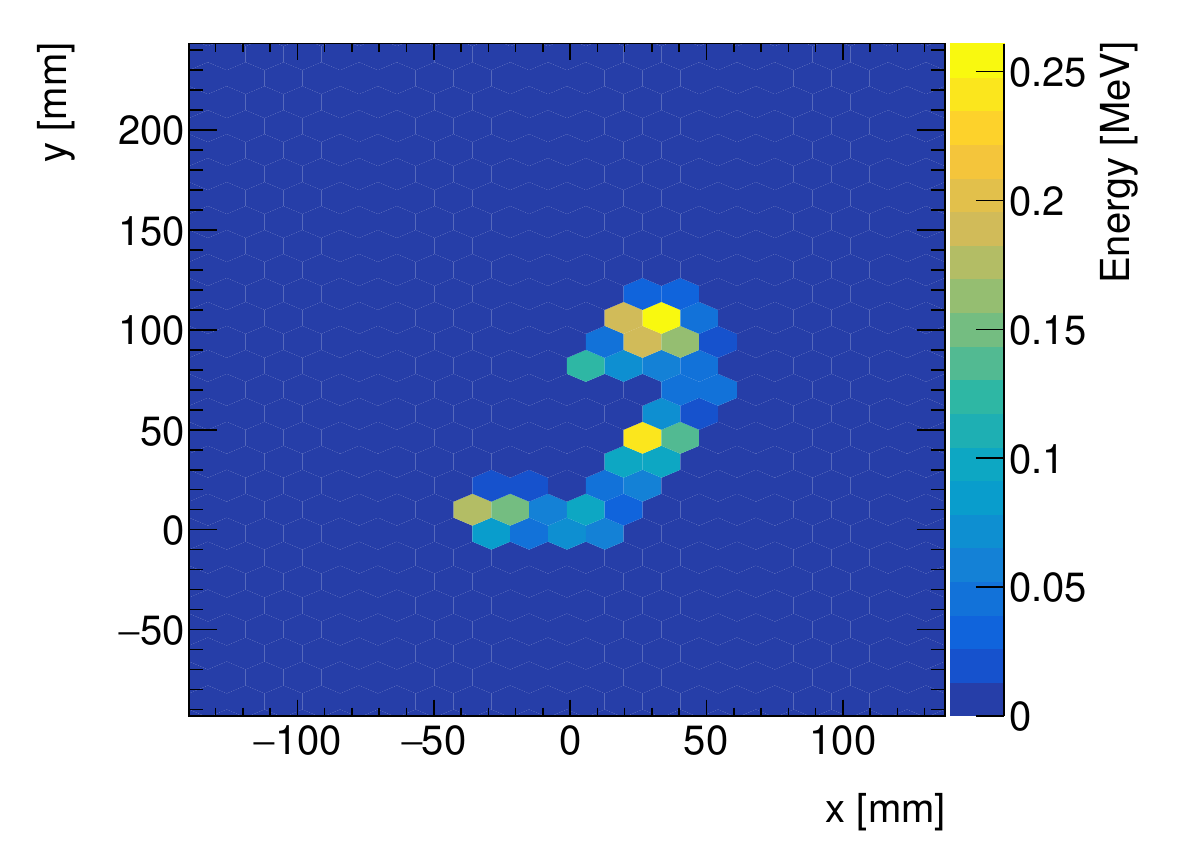}
    \includegraphics[width=0.48\linewidth]{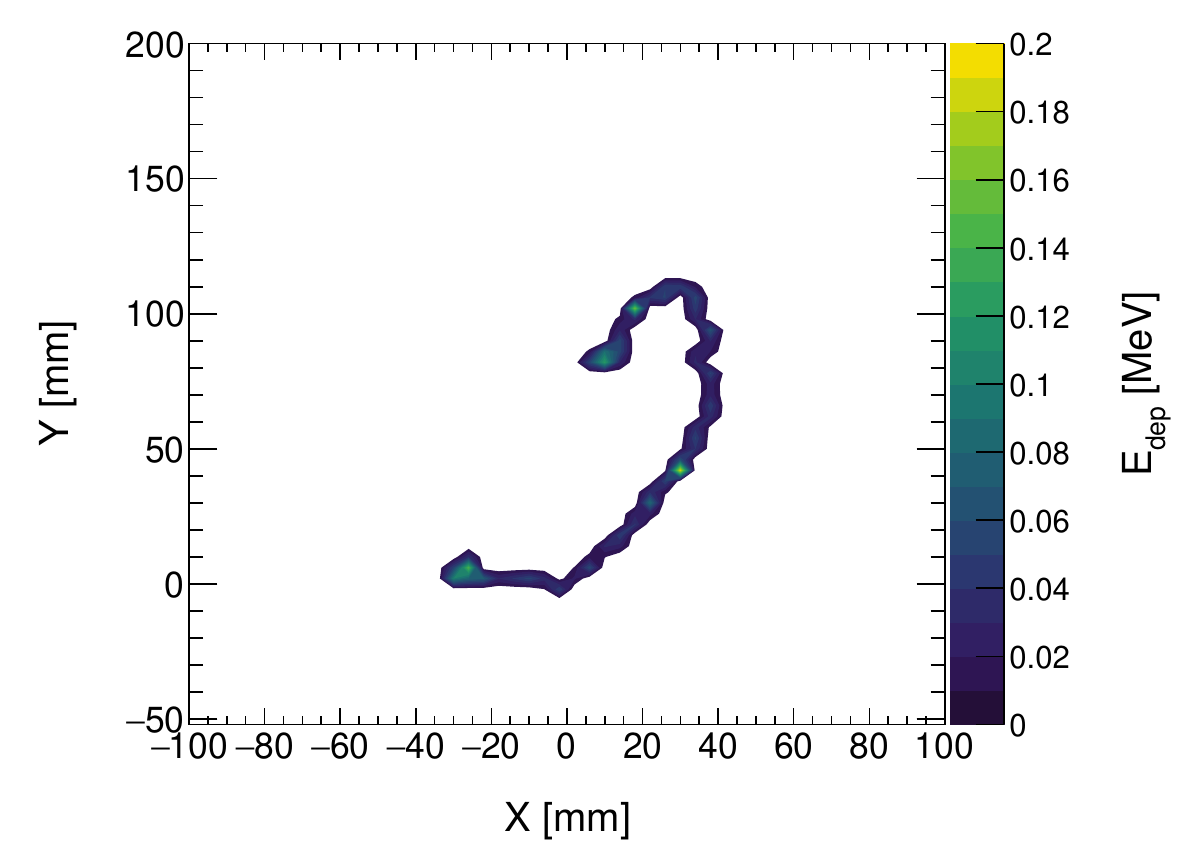}
    \caption{\reviewII{Calculated} energy distribution in the pixel array for a $0\nu\beta\beta$ event with a pixel noise of 40 e$^-$ (left), as well as the corresponding \Fix{true} energy depositions simulated by Geant4 (right).}
    \label{fig:PixvsG4}
\end{figure}

\rewrite{Garfield++ simulates the drift process using inputs from Geant4 (energy deposition) and COMSOL (field maps). The resulting induced currents are then converted into pixel voltage waveforms.}
Figure~\ref{fig:TimeDis} shows the \reviewII{simulated} distribution of drift times for all the charge carriers recorded at anode for the two ion species, whose existence was assumed. The ratio of mean drift times matches the inverse ratio of their mobilities.
Figure~\ref{fig:PixvsG4} shows the energy distribution in the pixel array for a $0\nu\beta\beta$ event, with a \reviewII{background} electronic noise of 40 e$^-$.
The corresponding Geant4 \Fix{true} energy depositions are also shown.

\begin{figure}
    \centering
    \includegraphics[width=0.9\linewidth]{ 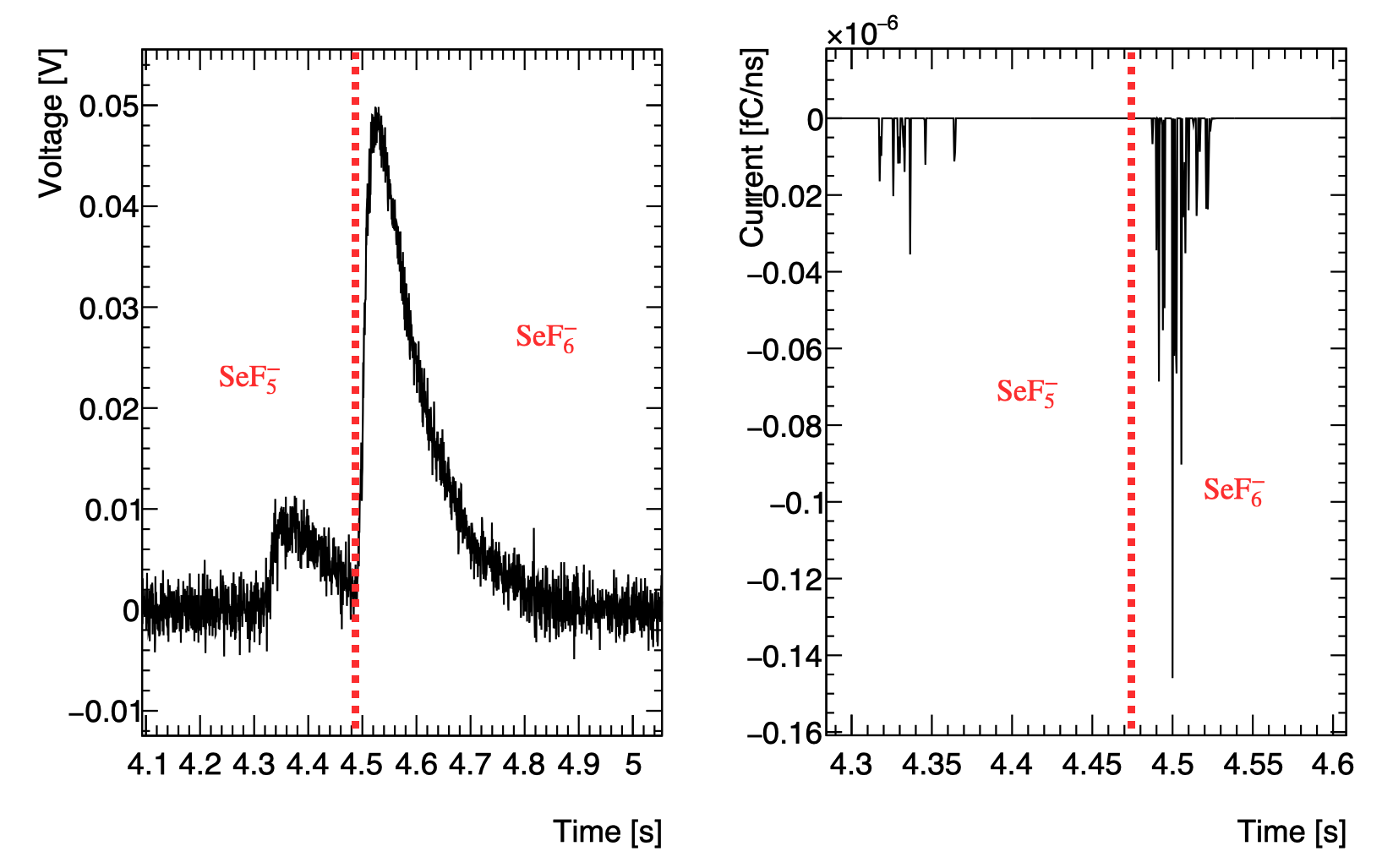}
    \caption{Simulated voltage (left) and induced current (right) waveforms for a pixel receiving both SeF$_5^-$ and SeF$_6^-$ ions. \reviewII{The red dashed line separates the two contributions.}. \reviewI{Despite different arrival times, the voltage signals overlap due to shaping effects.}}
    \label{fig:Sig}
\end{figure}

The energies per hit are reconstructed from the voltage waveform by convolving the induced current with the transfer function (Eq.~\ref{eq:Transfer}). To simulate the Topmetal-S response, we use fit parameters from Fig.~\ref{fig:WavCom} ($n = 0.2429$, $\tau = 7.14 \times 10^7$, $g = -2.837$).
Gaussian white noise (equivalent noise charge (ENC) $=$ 40 e$^-$) is added to the voltage waveform to reflect realistic conditions.
 The simulation uses a 10 s time window with a configurable sampling rate of 2 kS/s ($2\times10^4$ bins). Figure~\ref{fig:Sig} illustrates the resulting waveform contributions from both SeF$_5^-$ and SeF$_6^-$ ions (20:80 ratio) for the initial ionizations at z = 91.7 cm.
 Due to the long shaping time, individual contributions are not fully resolved in the voltage waveform.

\begin{figure}
    \centering
    \includegraphics[width=0.8\linewidth]{ 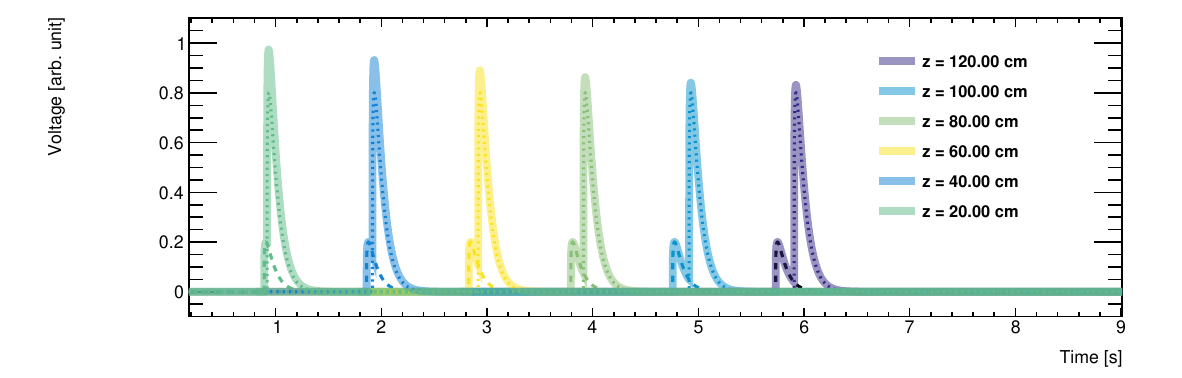}
    \caption{Waveform comparison for different drift lengths ($z$).
    Dashed lines show individual ion contributions; solid lines show the sum. Signals from longer drift distances are broader due to diffusion.}
    \label{fig:WavDiffZ}
\end{figure}

For sufficiently long drift lengths, the waveforms can be distinguished. In Fig~\ref{fig:WavDiffZ}, the waveforms under different drift lengths are compared. \reviewI{Because of longitudinal diffusion, the waveforms of signals from distant events are broadened, while signals near the anode ($z < 60$ cm) are difficult to separate.} The separation method for different drift lengths will be studied in future work.

\section{Analysis of track reconstruction}
\label{sec:analy}

\subsection{Topological reconstruction of 3D tracks}
\label{sec:Reco}

\reviewI{While a relative z-coordinate can be reconstructed using the arrival time of a single ion species, determining the absolute z-position is critical for rejecting surface backgrounds. Most radioactive backgrounds originate from hardware components at the boundaries of the drift volume. Therefore, defining a clean fiducial volume is a standard rejection strategy in high-pressure gas TPC experiments such as NEXT. For example, background studies from NEXT~\cite{NEXT:2019rum} demonstrate that a fiducial cut of 20 mm yields a selection efficiency of $(52.5 \pm 0.3)\%$ in data.}

\begin{figure}
    \centering
    \includegraphics[width=0.45\linewidth]{ 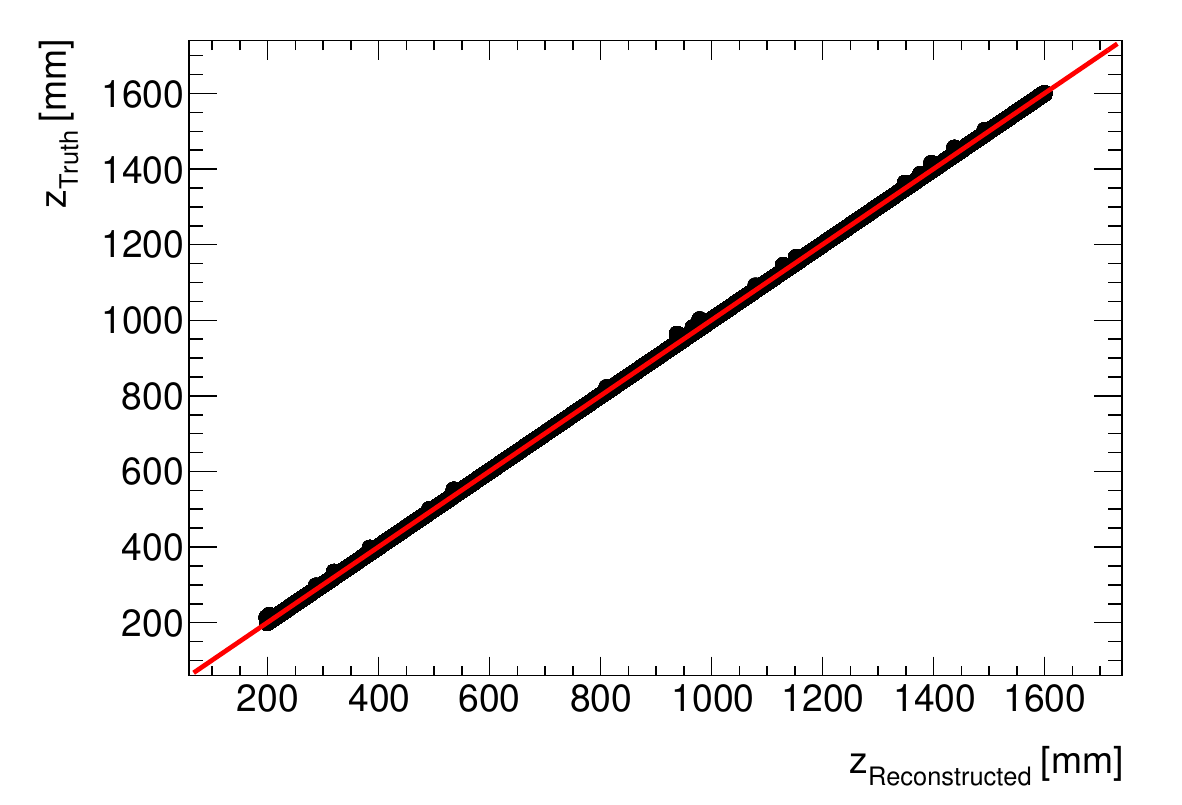}
    \caption{\rewrite{True versus reconstructed $z$ positions with a linear fit.} }
    \label{fig:ZComp}
\end{figure}

\reviewII{While the Topmetal-S sensor array intrinsically provides precise 2D (x-y) tracking information, full 3D reconstruction relies on the physical properties of the gas. Under the hypothesis that SeF$_6$ forms multiple negative ion species with distinct mobilities, the z-coordinate of each hit can be reconstructed from the drift time difference between the two assumed ion species:}
\begin{equation}
    z = \frac{v_{\mathrm{SeF_{5}^{-}}} \cdot v_{\mathrm{SeF_{6}^{-}}}}{v_{\mathrm{SeF_{5}^{-}}}- v_{\mathrm{SeF_{6}^{-}}}} \cdot \Delta t 
\end{equation}
where $\Delta t$ is the time difference between the \reviewII{arrival times of} of the two charge carriers at the anode. 
Figure~\ref{fig:ZComp} shows the comparison between the reconstructed and true z-coordinates. The gradient parameter of 1 is obtained from the linear fit, as expected.

\begin{figure}
    \centering
    \includegraphics[width=0.5\linewidth]{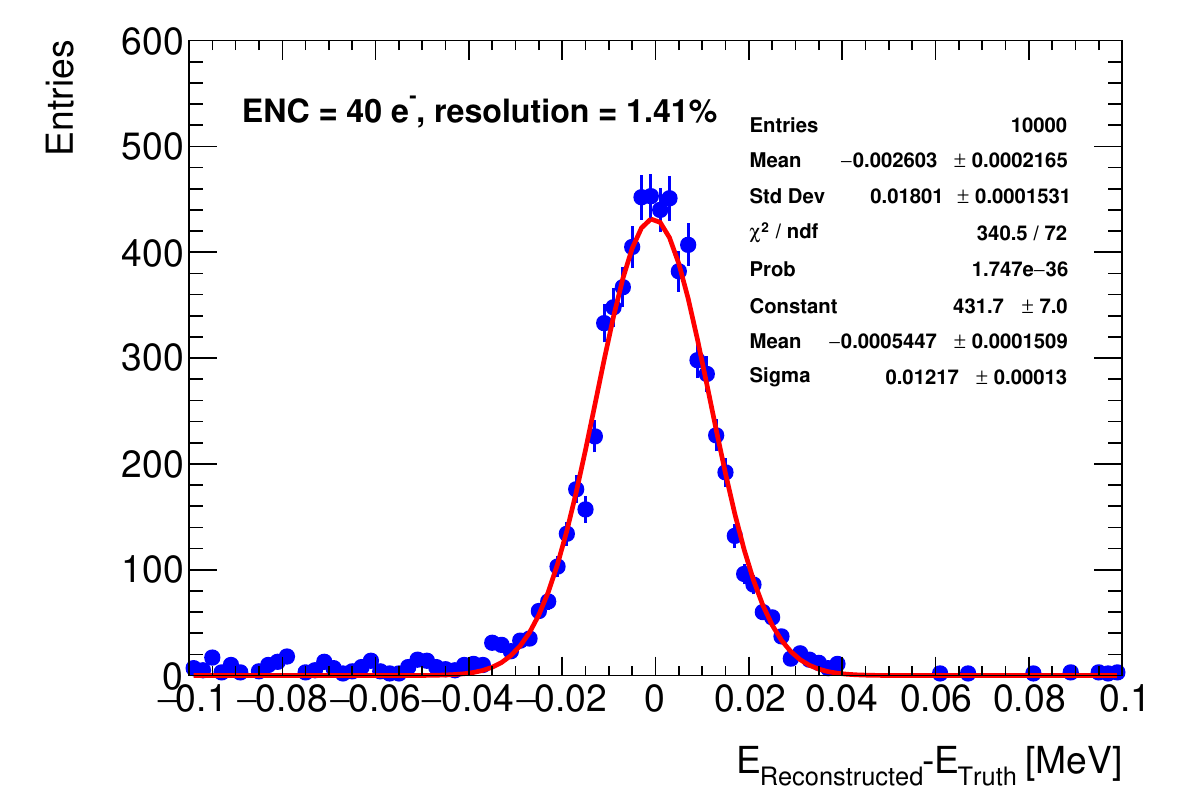}
    \caption{Energy resolution performance. The distribution shows the difference between reconstructed energy (with 40 e$^-$ ENC) and \Fix{true} energy. The energy resolution, defined as the FWHM of the distribution, is approximately 1.41\%. }
    \label{fig:Res}
\end{figure}

Each \rewrite{reconstructed} hit is treated as a node containing spatial (x, y, z) and energy information. To assemble these nodes into continuous tracks, we apply the Breadth-First Search (BFS) algorithm~\cite{cormen2001algorithms}, using a spatial adjacency criterion of 12 mm between neighboring nodes. This threshold is slightly larger than the pixel pitch (10 mm) \reviewII{to balance} reconstruction accuracy with computational efficiency.
The energy of each pixel was reconstructed from the total energy deposited within that pixel. Figure~\ref{fig:Res} presents the difference between reconstructed and true energies. When an ENC of 40 e$^-$ per hit pixel is included, the overall energy resolution (FWHM) is approximately \rewrite{1.41\%}.

\begin{figure}
    \centering
    \includegraphics[width=0.8\linewidth]{ 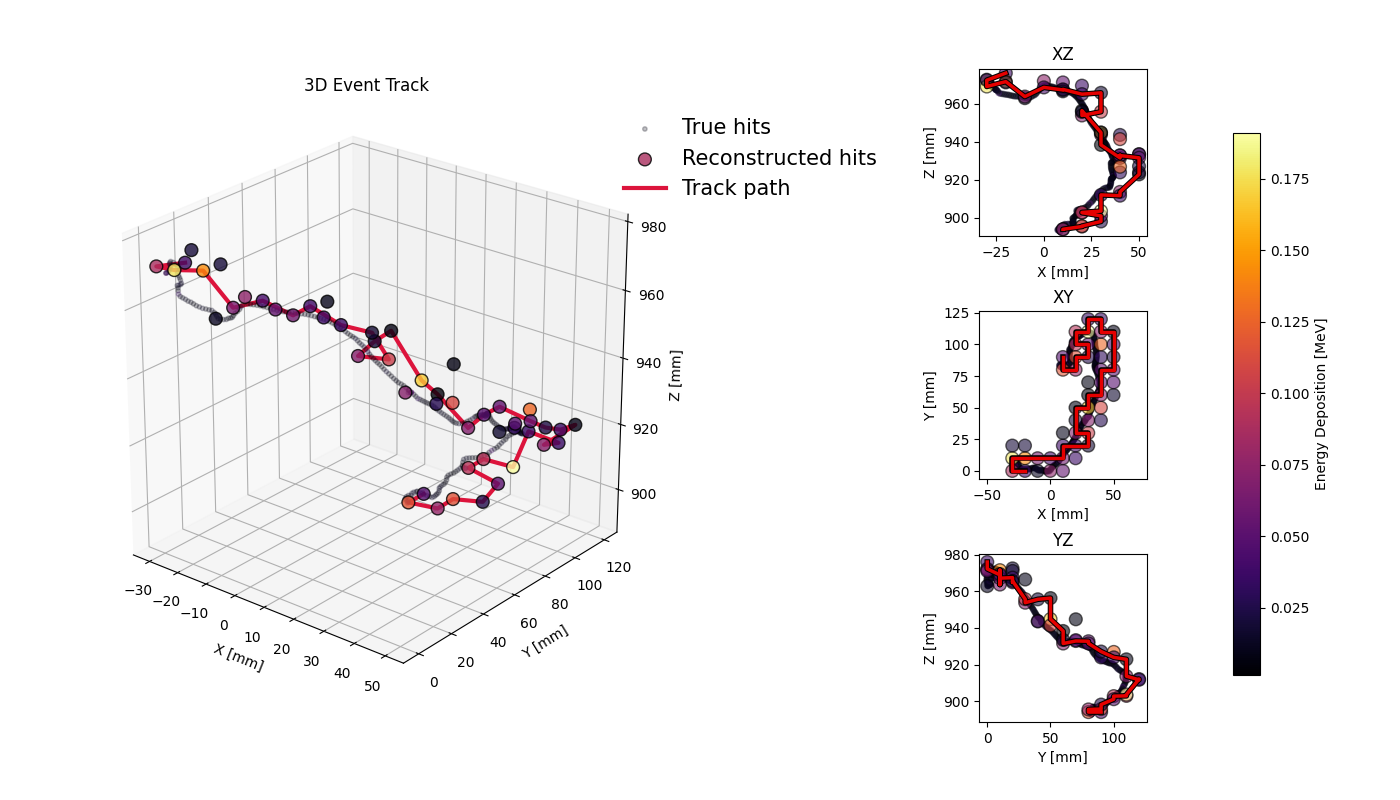}
    \caption{3D track reconstruction for a simulated $^{82}$Se $0\nu\beta\beta$ event.
    \reviewI{The plot displays both the reconstructed track (red path identified by BFS) and the corresponding \Fix{true} energy deposition, along with their 2D projections.}}
    \label{fig:3Dtrack}
\end{figure}

In cases where multiple disconnected graphs are formed, the one with the largest total energy deposition is selected as the main track, corresponding to the physical electron trajectory.
Figure~\ref{fig:3Dtrack} shows a reconstructed 3D track for a simulated $^{82}$Se $0\nu\beta\beta$ event with projections onto the x–y, x–z, and y–z planes. Two blobs, corresponding to the Bragg peaks of the emitted electrons, are expected for true $0\nu\beta\beta$ decays. The energy of blob is calculated by summing the energies within 50 mm of each main-track endpoint. 

\begin{figure}
    \centering
    \includegraphics[width=0.48\linewidth]{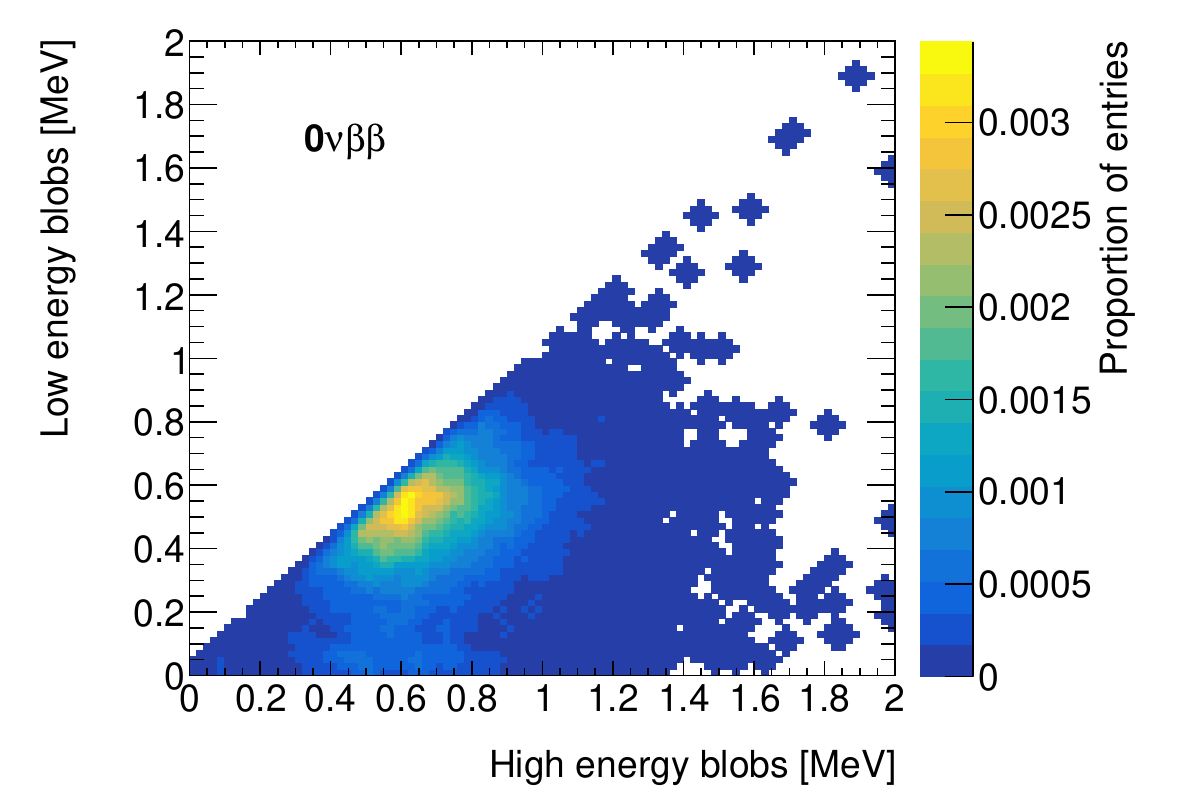}
    \includegraphics[width=0.48\linewidth]{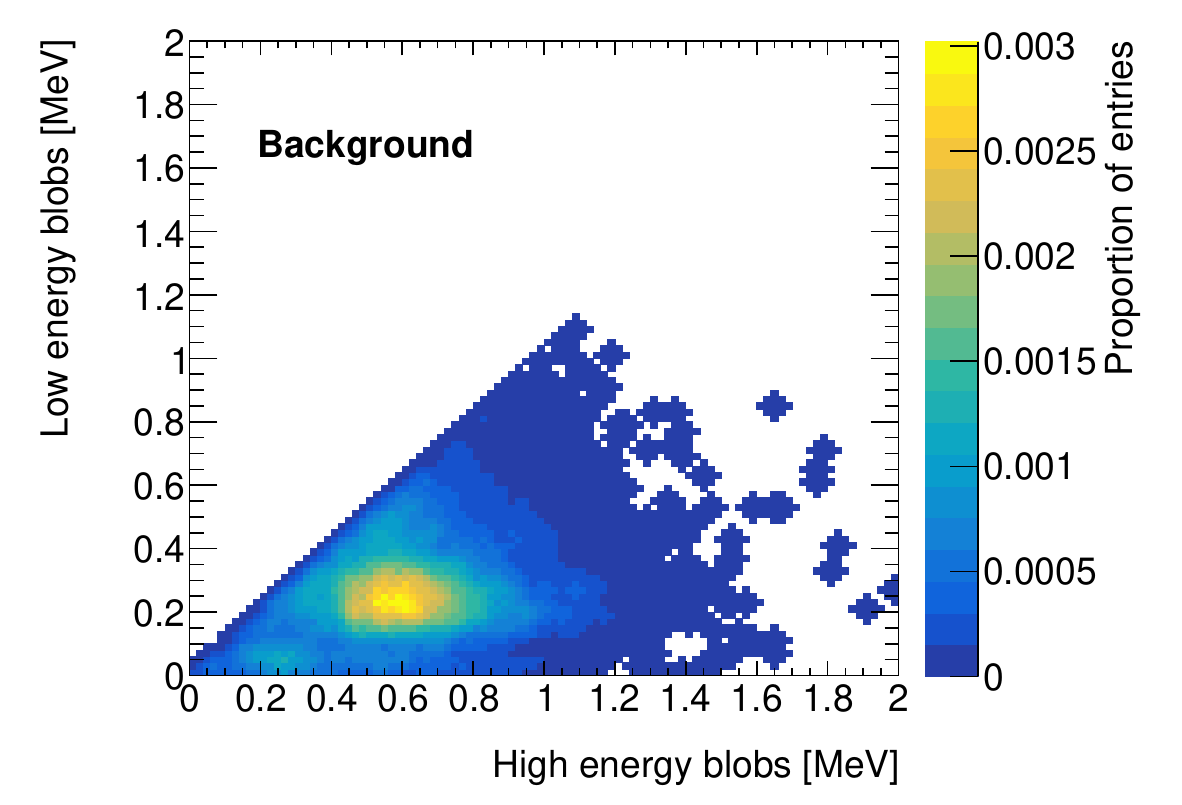}
    \caption{Energy deposition distribution for the two blob candidates. Left: $0\nu\beta\beta$ signal events showing two high-energy blobs. Right: Single-electron background with 2.998 MeV energy, typically showing one high-energy blob and one low-energy tail.}
    \label{fig:BlobDis}
\end{figure}

Blob identification is essential for discriminating $0\nu\beta\beta$ signal events from single-electron backgrounds originating from radioactive decays. Figure~\ref{fig:BlobDis} shows the energy distributions of blob candidates for signal and single-electron background with 2.998 MeV energy. Signal events typically exhibit two blobs with similar energies, while background events display a boarder and lower energy for low energy blobs.

\subsection{Preliminary background discrimination}
\label{sec:Discrimination}

Background events are primarily driven by single electrons from $\gamma$-ray interactions within the region of interest (ROI) and irreducible $2\nu\beta\beta$ decays.
With adequate tracking resolution, these can be distinguished from signal events through topological reconstruction.

\begin{figure}[h!]
    \centering
    \includegraphics[width=0.7\linewidth]{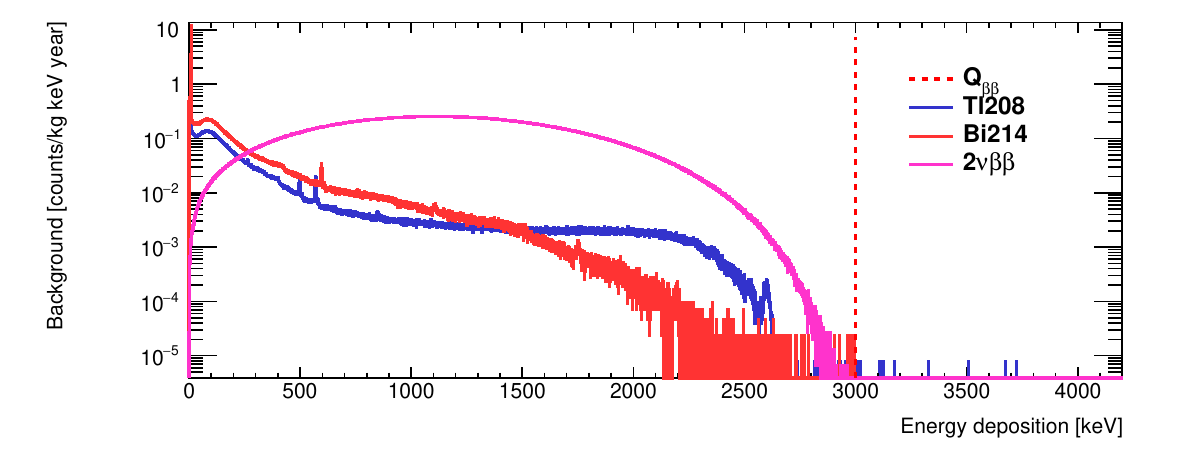}
    \caption{\rewrite{Simulated energy deposition spectra for $2\nu\beta\beta$ decay and dominant $\gamma$-ray backgrounds ($^{208}$Tl and $^{214}$Bi) in a 10 atm SeF$6$ gas.
    The vertical dashed line marks the Q-value of the $0\nu\beta\beta$ decay. }}
    \label{fig:GammaSpec}
\end{figure}

Figure~\ref{fig:GammaSpec} shows the simulated $\gamma$-ray spectra of $^{208}$Tl and $^{214}$Bi in 10 atm $^{82}$SeF$_6$. Both isotopes emit high-energy $\gamma$ rays, and could mimic the signal if the total energy deposited via Compton scattering or photoelectric absorption is within the ROI.
\rewrite{To normalize the radioactive $\gamma$ background from the copper shield, activity levels were referenced from the NEXT technical design report~\cite{Alvarez_2012}, assuming a total copper shield mass of 7584 kg. The normalized factor of 2$\nu\beta\beta$ is calculated using the $^{82}$Se half-life of $1.08 \times 10^{20}$ years (68\% C.L.)~\cite{Elliott:1992cf}.} 
$2\nu \beta \beta$ decays also contribute to the background by producing two electrons, \reviewII{although} these events are spectrally distinct. Since the neutrinos carry away most of the decay energy, the total deposited energy is generally well below the $0\nu\beta\beta$ $Q$-value of 2.998 MeV.
\rewrite{However, when finite energy resolution is considered, the high-energy tail of the $2\nu\beta\beta$ spectrum can \Fix{overlap} the ROI, as shown in Fig.~\ref{fig:GammaSpec}.}

 To study signal–background separation for the aforementioned $\gamma$ background, six topological variables are extracted from the reconstructed tracks from $0\nu\beta\beta$ signal and single-electron background, both with total kinetic energy of 2.998 MeV. Table.~\ref{tab:sumParameters} summarizes these variables.

\begin{table}[ht]
    \centering
    \begin{tabular}{c|c}
        \hline
        Symbol [Unit] & Meaning  \\
        \hline
        $L$ [mm]  & Main track length  \\
        $N$ [count]  & Number of hits \\
        $E$ [MeV]  & Total energy deposition \\
        $BL$ [MeV] & Low energy blob  \\
        $BH$ [MeV] & High energy blob \\
        $\mathrm{E_p}$ [MeV] & Total energy of main track \\
        \hline
    \end{tabular}
    \caption{Summary of the six topological variables used to distinguish $0\nu\beta\beta$ signals from single-electron backgrounds.}
    \label{tab:sumParameters}
\end{table}

\begin{figure}[h!]
    \centering
    \includegraphics[width=0.25\linewidth]{ 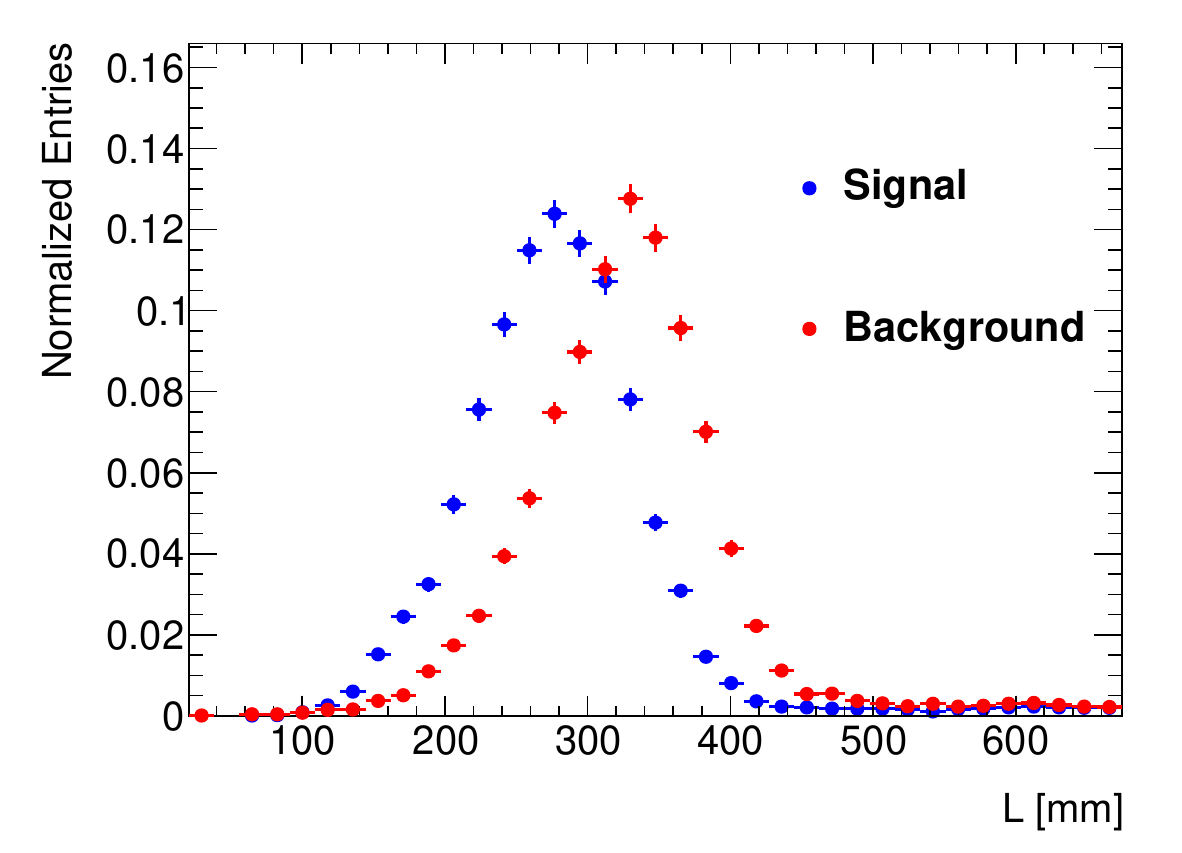}
    \includegraphics[width=0.25\linewidth]{ 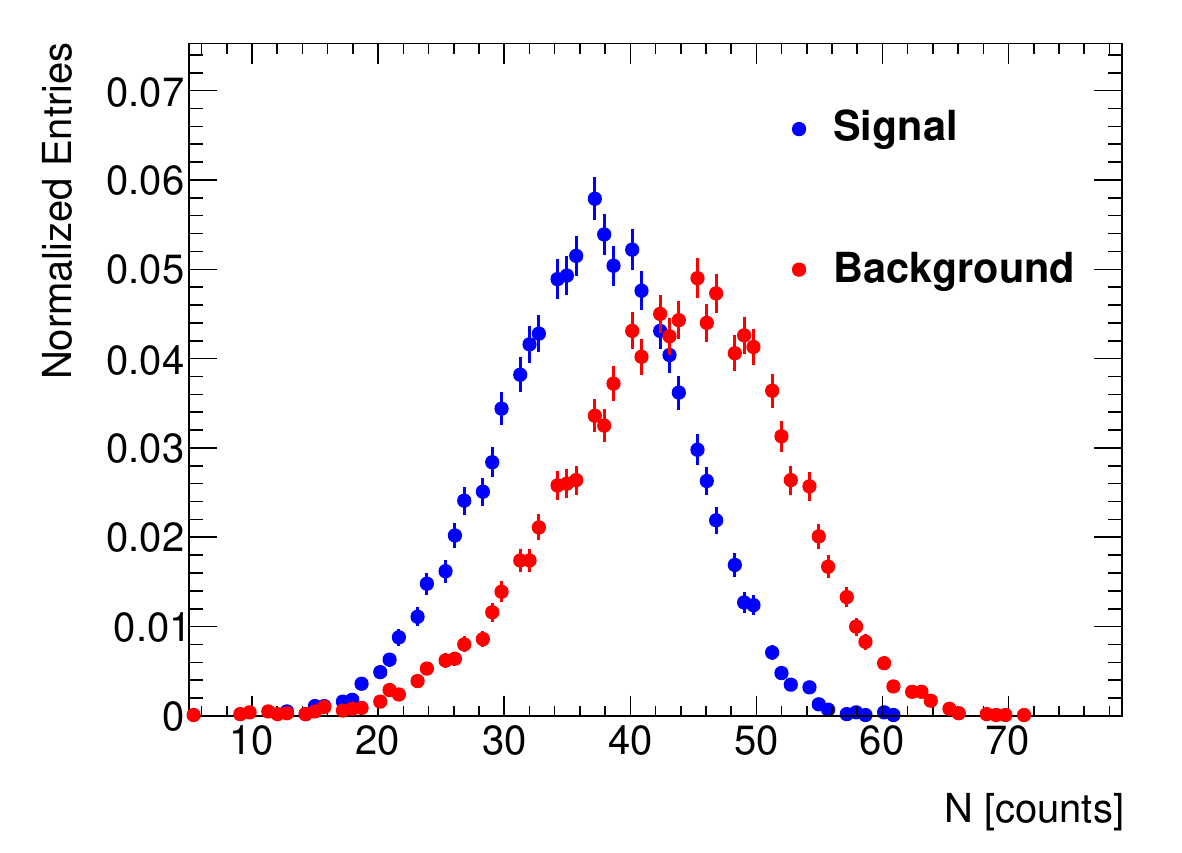}
    \includegraphics[width=0.25\linewidth]{ 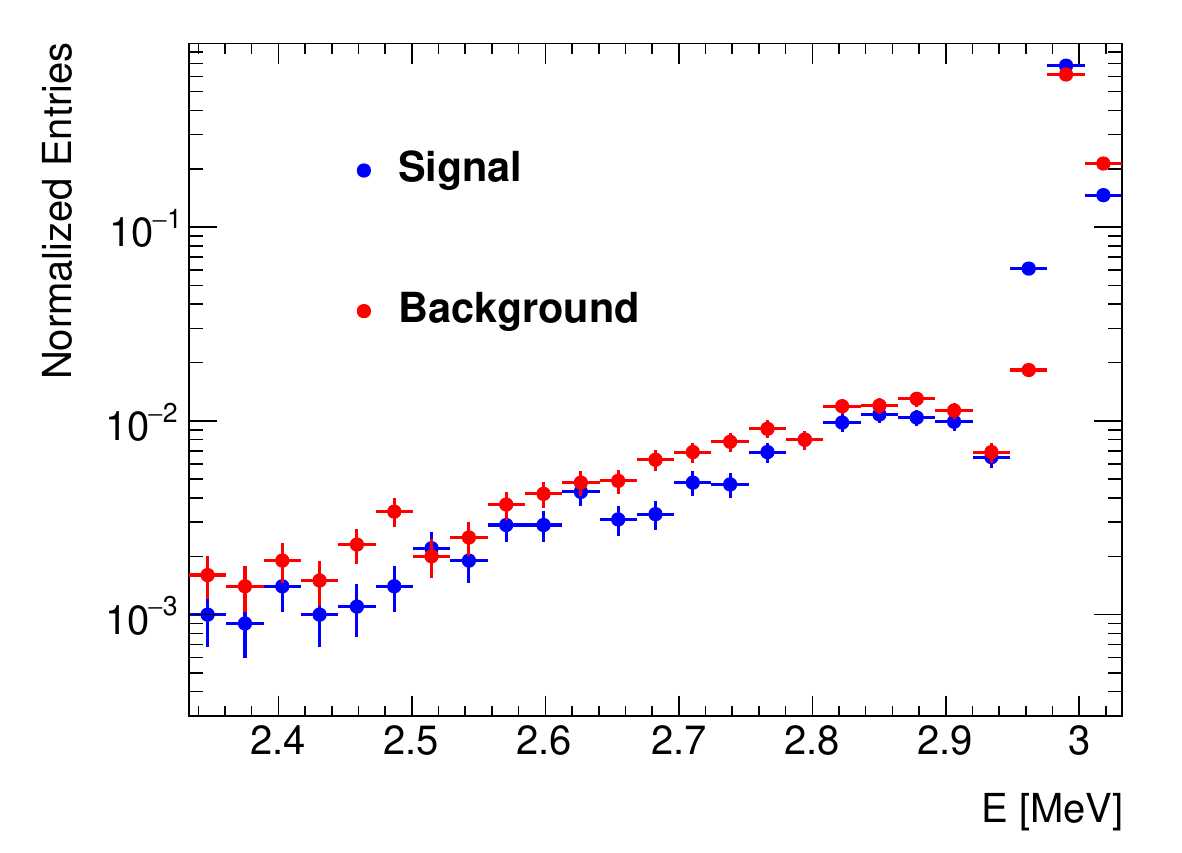}
    \includegraphics[width=0.25\linewidth]{ 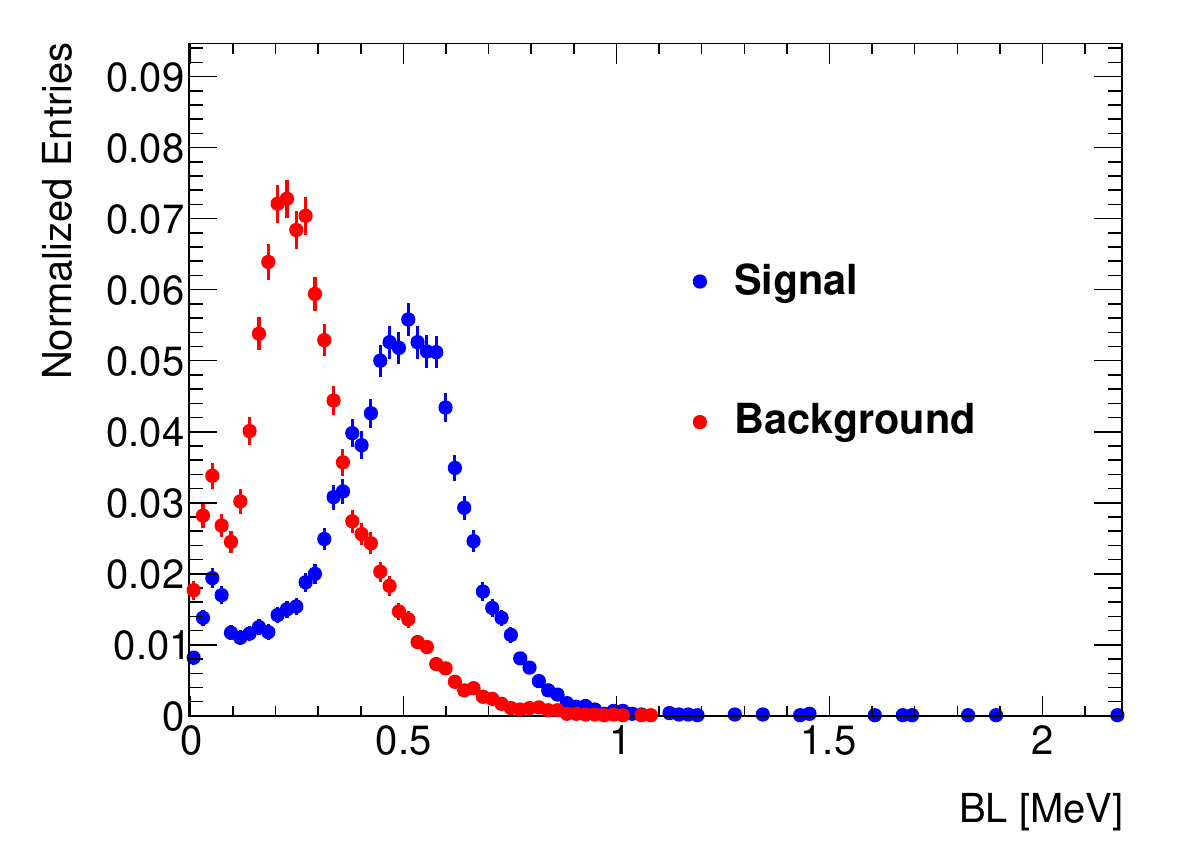}
    \includegraphics[width=0.25\linewidth]{ 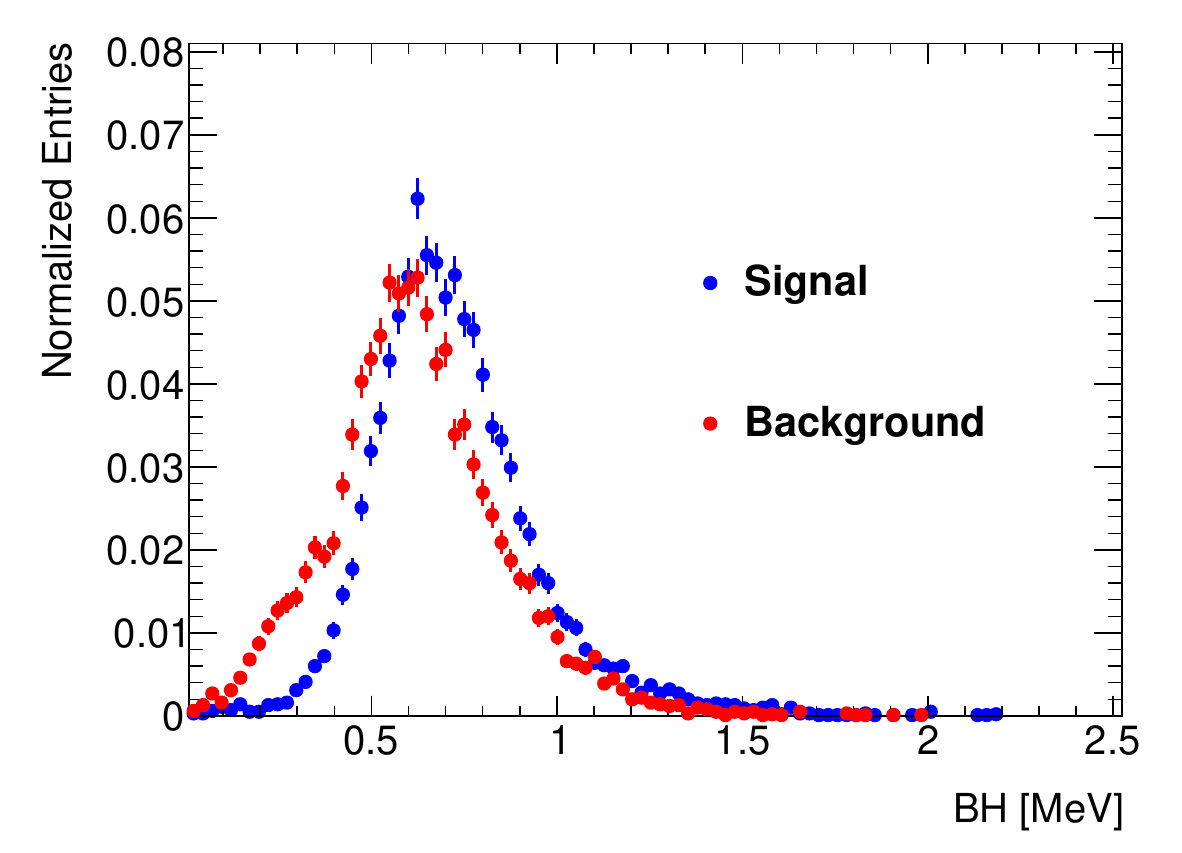}
    \includegraphics[width=0.25\linewidth]{ 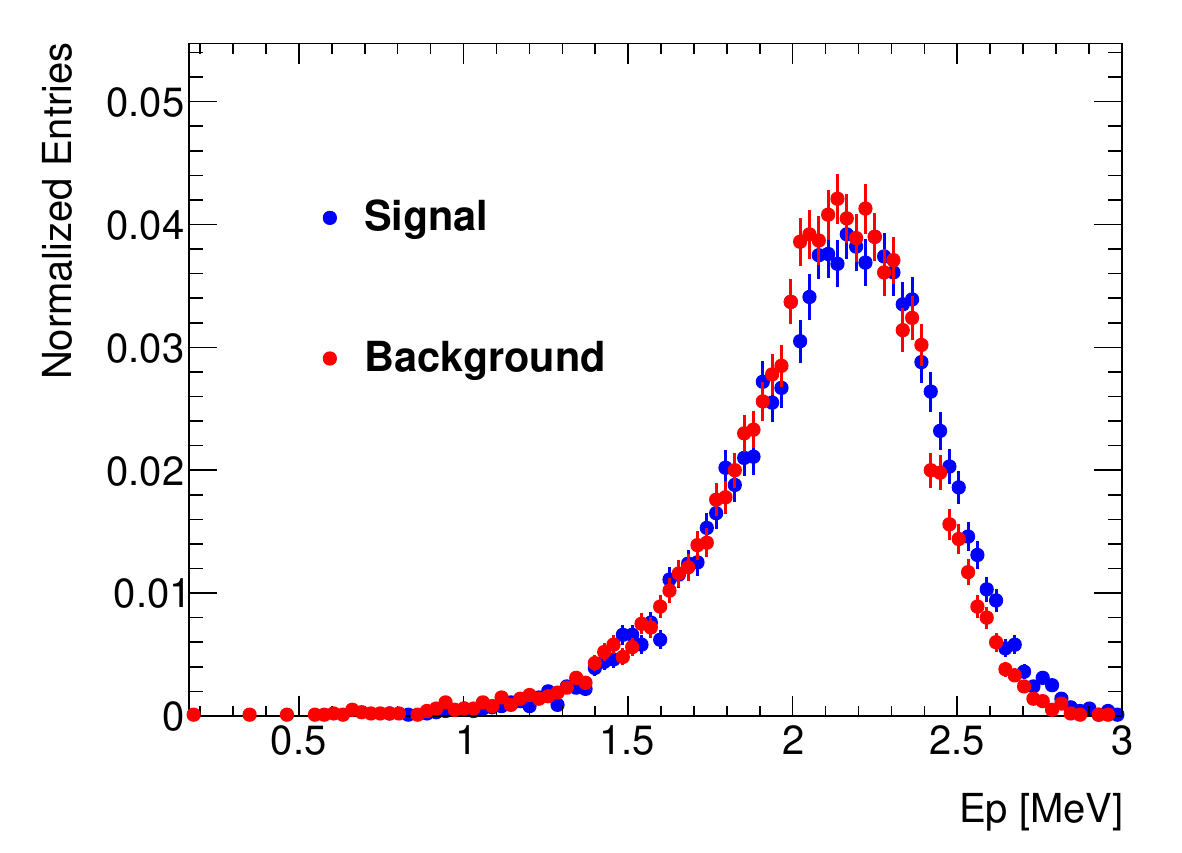}
    \caption{Distributions of the six topological variables for signal ($0\nu\beta\beta$) and single-electron background events, both with total kinetic energy of 2.998 MeV.}
    \label{fig:PareDis}
\end{figure}

\begin{figure}[h]
    \centering
    \includegraphics[width=0.4\linewidth]{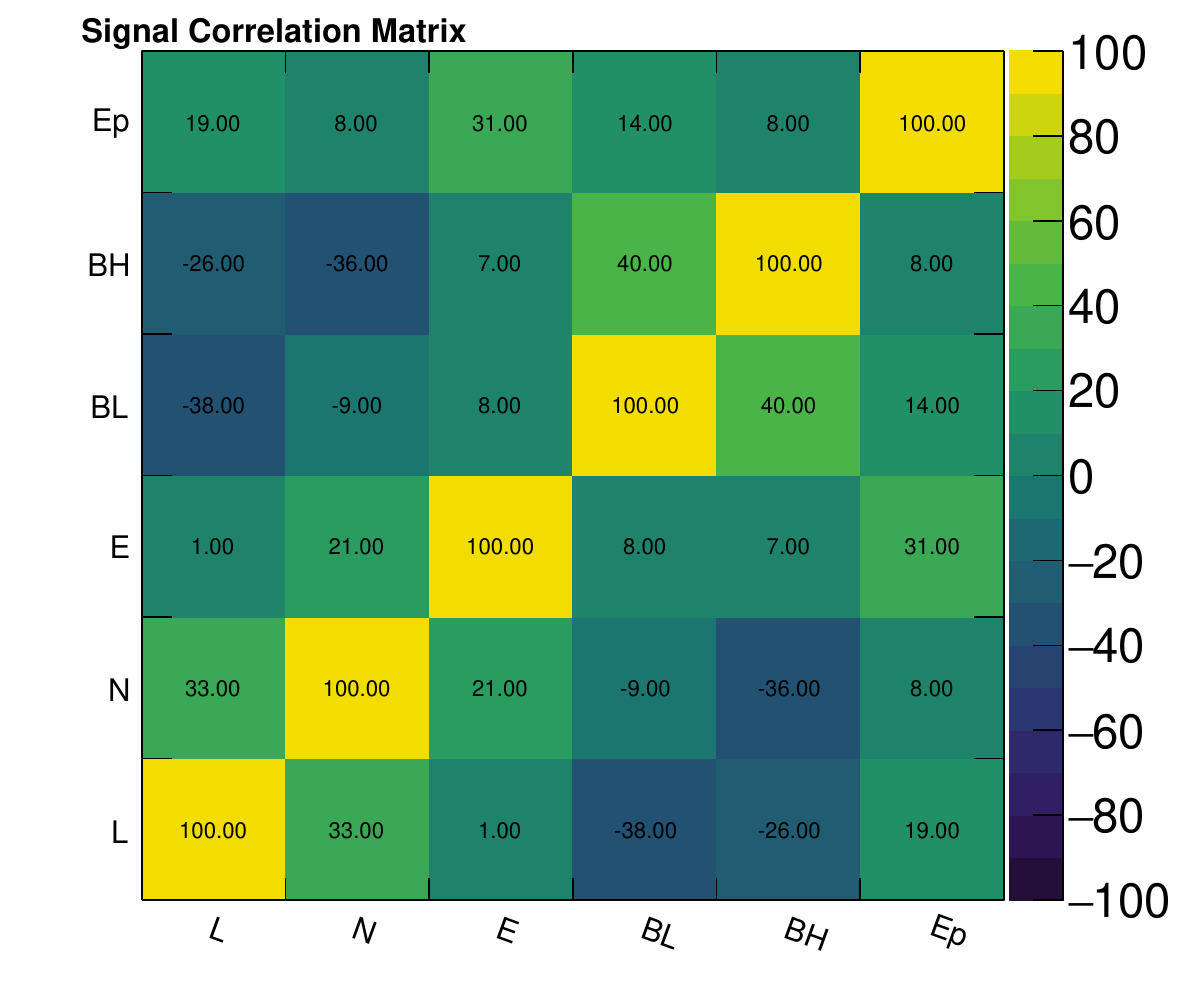}
    \includegraphics[width=0.4\linewidth]{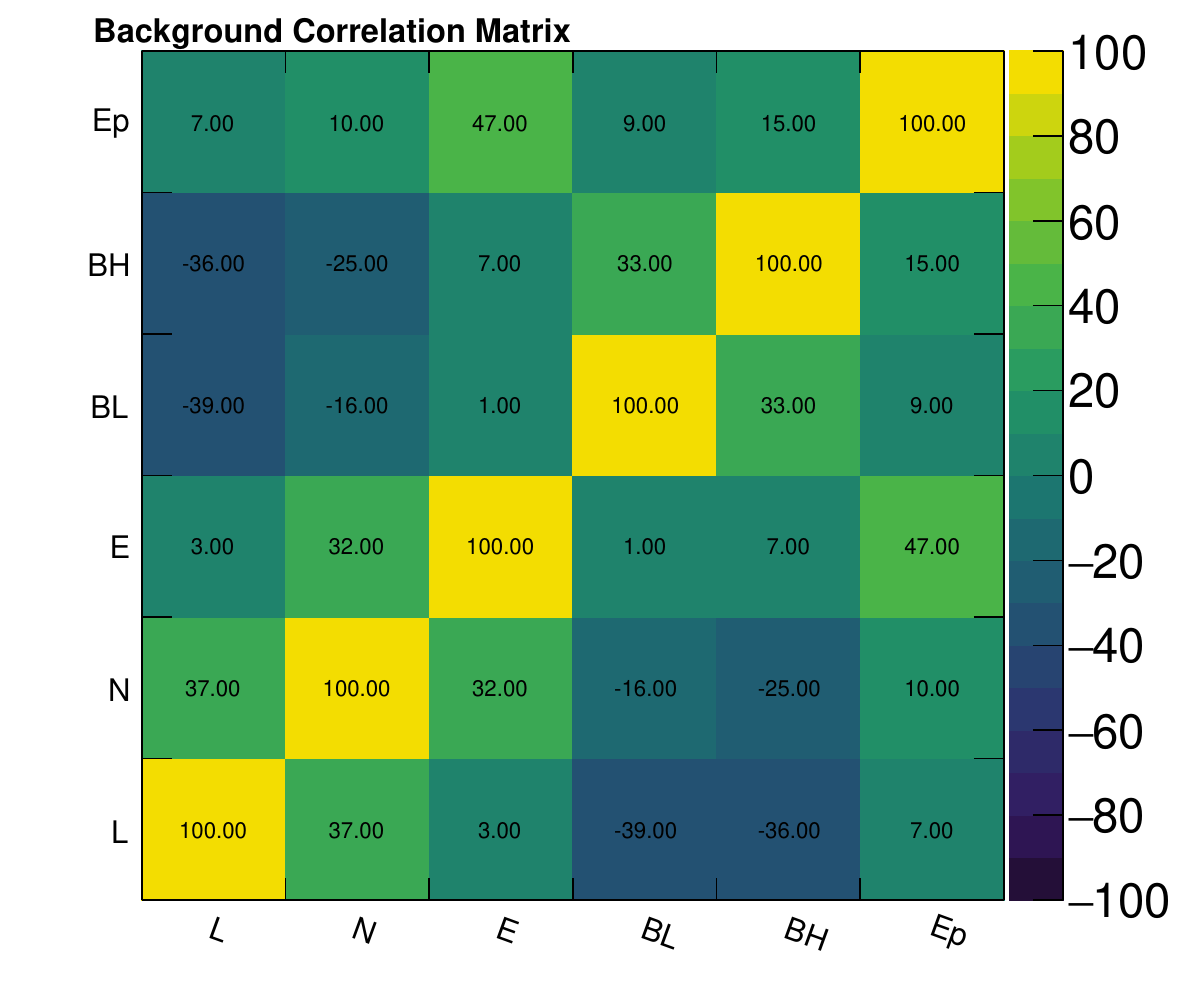}
    \caption{Correlation matrices of topological variables for signal (left) and single-electron background (right) events, both with total kinetic energy of 2.998 MeV.}
    \label{fig:CM}
\end{figure}

Figure~\ref{fig:PareDis} shows the distributions of \reviewII{the} six topological variables for $0\nu\beta\beta$ and background events. 
The correlation matrices in Fig.~\ref{fig:CM} display the relationship between these parameters. The single-electron backgrounds of 2.998 MeV tend to have longer track length and lower energy for low energy blob.

\begin{figure}[h!]
    \centering
    \includegraphics[width=0.45\linewidth]{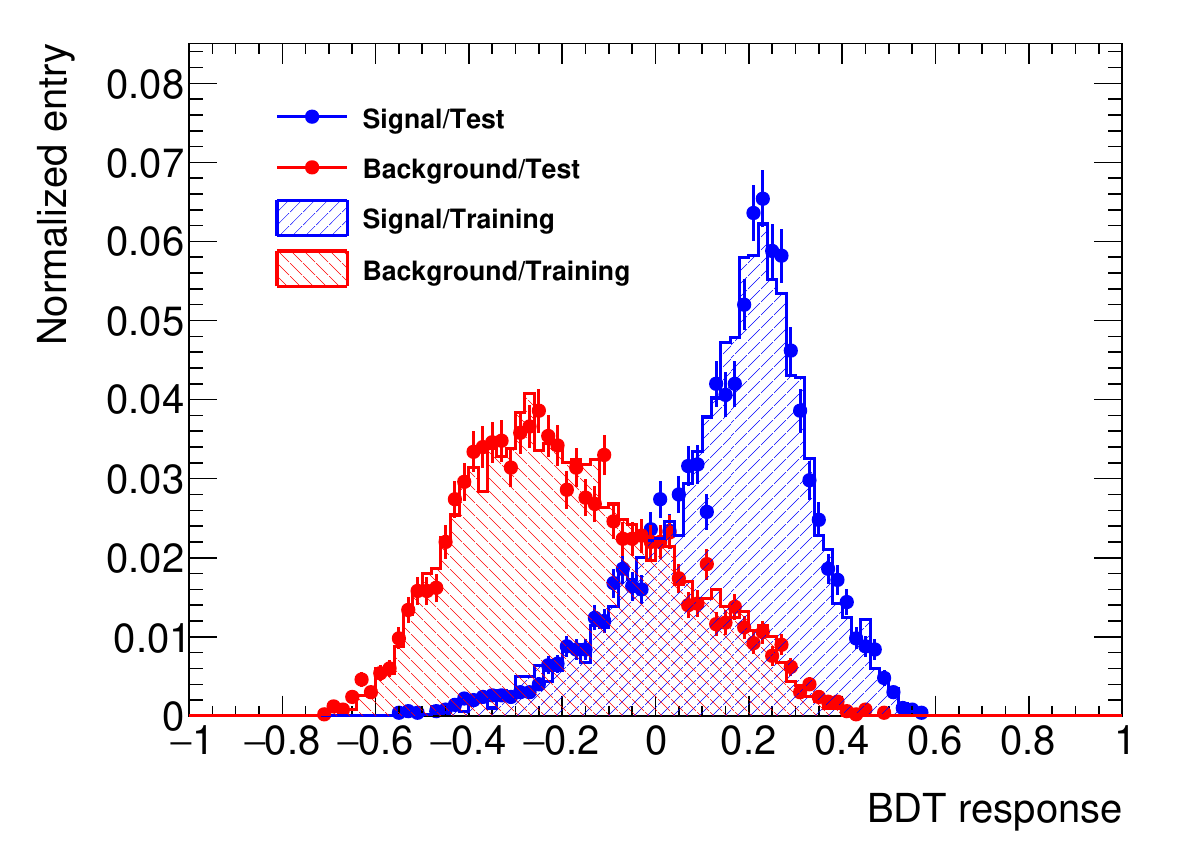}
    \includegraphics[width=0.45\linewidth]{ 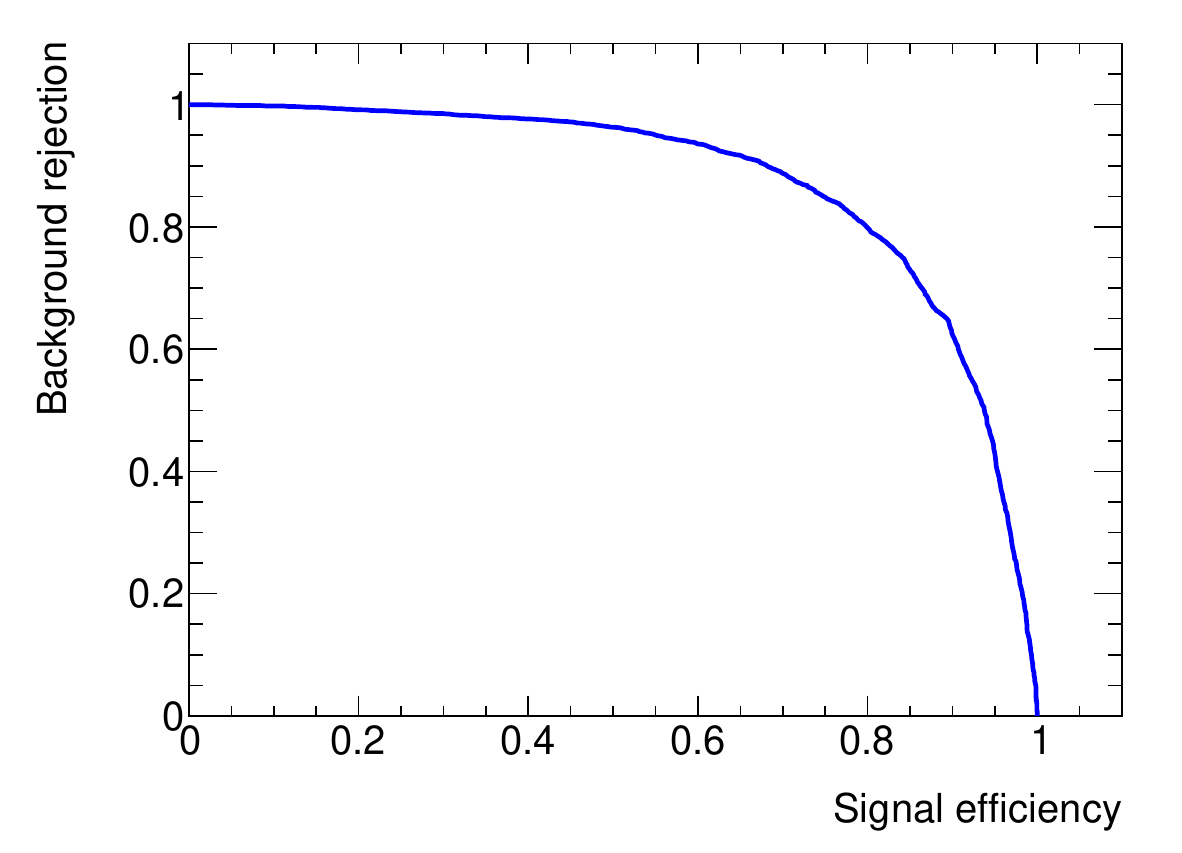}
    \caption{Performance of the BDT classifier. Left: BDT response distributions for signal and single-electron background samples (training vs. test), both with total kinetic energy of 2.998 MeV. Right: background rejection versus signal efficiency.}
    \label{fig:BDTCut}
\end{figure}

To combine the background discrimination capabilities of these variables, we applied a Boosted Decision Tree (BDT) analysis using TMVA~\cite{Hocker:2007ht}. Figure~\ref{fig:BDTCut} shows the BDT performance. The left plot illustrates the BDT response distributions for signal and background, and the right plot depicts the background rejection as a function of signal efficiency. 
For a signal efficiency of 75\% (90\%), the background rejection factor is 84\% (68\%).


\section{Summary}
\label{sec:Sum}

We have developed a simulation and analysis framework for the N$\nu$DEx experiment, which integrates molecular modeling, detector simulation and responses, and event reconstruction, providing a realistic evaluation of detector performance.

Negative ion formation in SeF$_6$ was \reviewII{tentatively simulated} by optimizing molecular geometries and calculating the mobilities of SeF$_5^-$ and SeF$_6^-$ \reviewII{ions}  using density functional theory and two-temperature theory, yielding charge-carrier drift velocities.

The detector geometry, featuring approximately 10,000 Topmetal-S pixel sensors, was modeled in COMSOL to obtain accurate electric field and weighting potential maps. Event generation and transport were simulated using BxDecay0 and Geant4, with Garfield++ employed for ion drift and signal induction. The sensor response was obtained by convolving the induced current with \Fix{the} transfer function, including realistic electronic noise.

\rewrite{Three-dimensional tracks were reconstructed using a BFS algorithm. Blob-finding techniques were implemented to distinguish signal-like double-electron tracks from single-electron background events. A preliminary analysis using a BDT classifier based on reconstructed tracks was performed within this framework, demonstrating its capability for signal–background discrimination.}

Future work will include the refinement of gas parameters through dedicated experimental measurements, \reviewII{optimization of} the electric field design, as well as \reviewII{refinement of} the algorithm on tracking and signal-background discrimination algorithms. 

\section*{Acknowledgment}

This work was supported in part by the National Key Research and Development Program of China under Grant 2022YFA1604703, 2021YFA1601300
and in part by the National Natural Science Foundation of China under Grant 12105110.

\bibliographystyle{elsarticle-num.bst}
\bibliography{biblio}

\end{document}